\documentclass{aa}

\usepackage{graphicx}
\usepackage{txfonts}
\usepackage{xcolor}

\newcommand{\adjustedaccent}[1]{%
  \mathchoice{}{}
    {\mbox{\raisebox{-.5ex}[0pt][0pt]{$\scriptstyle#1$}}}
    {\mbox{\raisebox{-.35ex}[0pt][0pt]{$\scriptscriptstyle#1$}}}
}
\newcommand\vv[1]{\overset{\adjustedaccent{\rightharpoonup}}{#1}}

\makeatletter
\newcommand{\oset}[3][0ex]{%
  \mathrel{\mathop{#3}\limits^{
    \vbox to#1{\kern-1\ex@
    \hbox{$\scriptstyle#2$}\vss}}}}
\makeatother

\newcommand\TrigFlux[1]{\mathcal{G}^{#1}}
\newcommand\DetrendedFlux{F}
\newcommand\ParamVec{\vv{\theta}}
\newcommand\Model[1]{\mathcal{M}_{#1}}
\newcommand\NoMoonModel{\mathcal{M}_0}
\newcommand\OneMoonModel{\mathcal{M}_1}
\newcommand\ModelFlux[1]{\mathcal{F}_{#1}}

\usepackage[pdftitle={Revisiting the exomoon candidate signal around Kepler-1625\,b}, colorlinks = true, breaklinks = true, citecolor = blue, linkcolor = blue, urlcolor = blue, pdfauthor = {Kai Rodenbeck}]{hyperref}

\begin{document} 

\title{Revisiting the exomoon candidate signal around Kepler-1625\,b}

\author{Kai Rodenbeck \inst{1,2}
             \and
             Ren\'{e} Heller\inst{2}
             \and
             Michael Hippke \inst{3}
             \and
             Laurent Gizon \inst{2,1}
}

\institute{Institute for Astrophysics, Georg August University G\"ottingen, Friedrich-Hund-Platz 1, 37077 G\"ottingen, Germany
           \and 
           Max Planck Institute for Solar System Research, Justus-von-Liebig-Weg 3, 37077 G\"ottingen, Germany \\
           \email{(rodenbeck/heller/gizon)@mps.mpg.de}
           \and
           Sonneberg Observatory, Sternwartestr. 32, 96515 Sonneberg, Germany \email{michael@hippke.org}
}

   \date{Accepted 26.6.2018}
 
  \abstract
   { Transit photometry of the Jupiter-sized exoplanet candidate Kepler-1625\,b has recently been interpreted to show hints of a moon. This exomoon, the first of its kind, would be as large as Neptune and unlike any moon we know from the solar system.}
   { We aim to clarify whether the exomoon-like signal is indeed caused by a large object in orbit around Kepler-1625\,b, or whether it is caused by stellar or instrumental noise or by the data detrending procedure. }
   {
   To prepare the transit data for model fitting, we explore several detrending procedures using second-, third-, and fourth-order polynomials and an implementation of the Cosine Filtering with Autocorrelation Minimization (CoFiAM).
   We then supply a light curve simulator with the co-planar orbital dynamics of the system and fit the resulting planet-moon transit light curves to the Kepler data. 
   We employ the Bayesian Information Criterion (BIC) to assess whether a single planet or a planet-moon system is a more likely interpretation of the light curve variations. 
   We carry out a blind hare-and-hounds exercise using many noise realizations by injecting simulated transits into different out-of-transit parts of the original Kepler-1625 light curve: (1) 100 sequences with three synthetic transits of a Kepler-1625\,b-like Jupiter-size planet and (2) 100 sequences with three synthetic transits of a Kepler-1625\,b-like planet with a Neptune-sized moon.
   }
   { The statistical significance and characteristics of the exomoon-like signal strongly depend on the detrending method (polynomials versus cosines), the data chosen for detrending, and on the treatment of gaps in the light curve. Our injection-retrieval experiment shows evidence of moons in about 10\,\% of those light curves that do not contain an injected moon. Strikingly, many of these false-positive moons resemble the exomoon candidate, i.e. a Neptune-sized moon at about 20 Jupiter radii from the planet. We recover between about a third and half of the injected moons, depending on the detrending method, with radii and orbital distances broadly  corresponding to the injected values.}
   {A ${\Delta}{\rm BIC}$ of $-4.9$ for the CoFiAM-based detrending is indicative of an exomoon in the three transits of Kepler-1625\,b. This solution, however, is only one out of many and we find very different solutions depending on the details of the detrending method. We find it concerning that the detrending is so clearly key to the exomoon interpretation of the available data of Kepler-1625\,b. Further high-accuracy transit observations may overcome the effects of red noise but the required amount of additional data might be large.}

   \keywords{ Planets and satellites: detection -- Eclipses -- Techniques: photometric -- Methods: data analysis }

   \maketitle
%

\newcommand{\degrees}{^\circ}

\graphicspath{{./figs/}}

\section{Introduction}
\begin{figure*}
\centering
\includegraphics[width=0.95\linewidth]{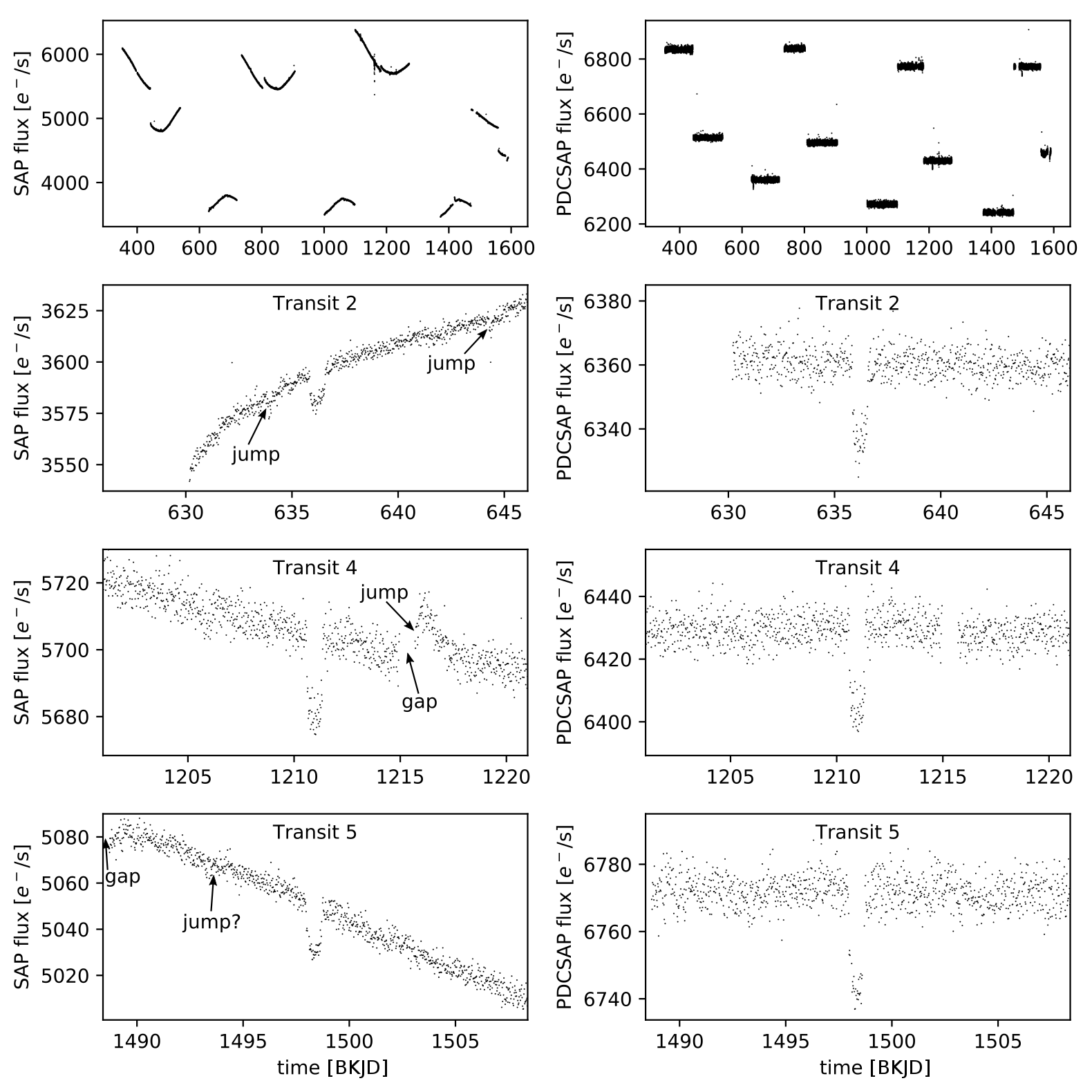}
\caption{Kepler light curve of Kepler-1625. \textit{Left:} Simple Aperture Photometry (SAP) flux. \textit{Right:} Pre-search Data Conditioning Simple Aperture Photometry (PDCSAP) flux. The top panels show the entire light curves, respectively. The second to fourth rows illustrate zooms into transits 2, 4, and 5 of Kepler-1625\,b, respectively. These transits were shifted to the panel center and $\pm10$\,d of data are shown around the transit mid-points. Some examples of jumps and gaps in the light curve are shown. Time is in Barycentric Kepler Julian Date.}
\label{fig:SAP_PDCSAP}
\end{figure*}
Where are they? -- With about 180 moons discovered around the eight solar system planets and over $3{,}500$ planets confirmed beyond the solar system, an exomoon detection could be imminent. While many methods have indeed been proposed to search for moons around extrasolar planets \citep{1999A&AS..134..553S,2002ApJ...580..490H,Cabrera2007,2009AsBio...9..269M,2009MNRAS.392..181K,2010MNRAS.406.2038S,Peters2013,2014ApJ...787...14H,BenJaffel2014,2015ApJ...812....5A,Forgan2017,2018arXiv180501903V}\footnote{For reviews see \citet{2014AsBio..14..798H} and \citet{HellerReview2017}.}, only a few dedicated surveys have actually been carried out \citep{2013A&A...553A..17S,2013ApJ...770..101K,2013ApJ...777..134K,2014ApJ...784...28K,2015ApJ...806...51H,2015ApJ...813...14K,2017A&A...603A.115L,2018AJ....155...36T}, one of which is the ``Hunt for Exomoons with Kepler'' \citep[HEK for short;][]{2012ApJ...750..115K}.

In the latest report of the HEK team, \cite{2018AJ....155...36T} find evidence for an exomoon candidate around the roughly Jupiter-sized exoplanet candidate Kepler-1625\,b, which they provisionally refer to as Kepler-1625\,b-i. Kepler-1625 is a slightly evolved G-type star with a mass of $M_\star~=~1.079_{-0.138}^{+0.100}\,M_\odot$ ($M_\odot$ being the solar mass), a radius of $R_\star~=~1.793_{-0.488}^{+0.263}\,R_\odot$ (with $R_\odot$ as the solar radius), and an effective temperature of $T_{{\rm eff},\star}~=~5548_{-72}^{+83}$\,K \citep{2017ApJS..229...30M}. Its Kepler magnitude of 15.756 makes it a relatively dim Kepler target.\footnote{NASA Exoplanet Archive: \href{https://exoplanetarchive.ipac.caltech.edu}{https://exoplanetarchive.ipac.caltech.edu}} The challenge of this tentative detection is in the noise properties of the data, which are affected by the systematic noise of the Kepler space telescope and by the astrophysical variability of the star. Although the exomoon signal did show up both around the ingress/egress regions of the phase-folded transits \citep[known as the orbital sampling effect;][]{2014ApJ...787...14H,2016ApJ...820...88H} generated by \cite{2018AJ....155...36T} and in the sequence of the three individual transits, it could easily have been produced by systematics or stellar variability, as pointed out in the discovery paper.

The noise properties also dictate a minimum size for an exomoon to be detected around a given star and with a given instrument. In the case of Kepler-1625 we calculate the root-mean-square of the noise level to be roughly 700~ppm. As a consequence, any moon would have to be at least about $\sqrt{700\,{\rm ppm}}~{\times}~1.793\,R_\odot\approx 5.2\,R_\oplus$  ($R_\oplus$ being the Earth's radius) in size, about 30\% larger than Neptune, in order to significantly overcome the noise floor in a single transit. The three observed transits lower this threshold by a factor of $\sqrt{3}$, suggesting a minimum moon radius of $\approx 3\,R_\oplus$. In fact, the proposed moon candidate is as large as Neptune, making this system distinct from any planet-moon system known in the solar system \citep{2018A&A...610A..39H}.

Here we present a detailed study of the three publicly available transits of Kepler-1625\,b. Our aim is to test whether the planet-with-moon hypothesis is favored over the planet-only hypothesis. In brief, we
\begin{enumerate}
\item develop a model to simulate photometric transits of a planet with a moon (see Sect.~\ref{sec:onemoon}).
\item implement a detrending method following \citet{2018AJ....155...36T} and explore alternative detrending functions.
\item detrend the original Kepler-1625 light curve, determine the most likely moon parameters, and assess if the planet-with-moon hypothesis is favored over the planet-only hypothesis.
\item perform a blind injection-retrieval test. To preserve the noise properties of the  Kepler-1625 light curve, we inject planet-with-moon and planet-only transits into out-of-transit parts of the Kepler-1625 light curve.
\end{enumerate}

\section{Methods}


The main challenge in fitting a parameterized, noise-less model to observed data is in removing noise on time scales similar or larger than the time scales of the effect to be searched; at the same time, the structure of the effect shall be untouched, an approach sometimes referred to as ``pre-whitening'' of the data \citep{2004MNRAS.350..331A}. The aim of this approach is to remove unwanted variations in the data, e.g. from stellar activity, systematics, or instrumental effects. This approach bears the risk of both removing actual signal from the data and of introducing new systematic variability. The discovery and refutal of the exoplanet interpretation of variability in the stellar radial velocities of $\alpha$\,Centauri\,B serves as a warning example \citep{2012Natur.491..207D,2016MNRAS.456L...6R}. Recently developed Gaussian process frameworks, in which the systematics are modeled simultaneously with stellar variability, would be an alternative method \citep{2012MNRAS.419.2683G}. This has become particularly important for the extended Kepler mission (K2) that is now working with degraded pointing accuracy \citep{2015MNRAS.447.2880A}.

That being said, \citet{2018AJ....155...36T} applied a pre-whitening technique to both the Simple Aperture Photometry (SAP) flux and the Pre-search Data Conditioning (PDCSAP) flux of Kepler-1625 to determine whether a planet-only or a planet-moon model is more likely to have caused the observed Kepler data. In the following, we develop a detrending and model fitting procedure that is based on the method applied by \cite{2018AJ....155...36T}, and then we test alternative detrending methods.

During Kepler's primary mission, the star Kepler-1625 has been monitored for 3.5 years in total, and five transits could have been observed. This sequence of transits can be labeled as transits 1, 2, 3, 4, and 5. Due to gaps in the data, however, only three transits have been covered, which correspond to transits 2, 4, and 5 in this sequence. Figure~\ref{fig:SAP_PDCSAP} shows the actual data set that we discuss. The entire SAP (left) and PDCSAP (right) light curves are shown in the top panels, and close-up inspections of the observed transit 2, 4, and 5 are shown in the remaining panels. The time system used throughout the article is the Barycentric Kepler Julian Date (BKJD), unless marked as relative to a transit midpoint.

\subsection{Detrending}
A key pitfall of any pre-whitening or detrending method is the unwanted removal of signal or injection of systematic noise, the latter of which could mimic signal. In our case of an exomoon search, we know that the putative signal would be restricted to a time-window around the planetary mid-transit, which is compatible with the orbital Hill stability of the moon. This criterion defines a possible window length that we should exclude from our detrending procedures. For a nominal 10 Jupiter-mass planet in a 287\,d orbit around a $1.1\,M_\odot$ star \citep[as per][]{2018AJ....155...36T}, this window is about 3.25 days to both sides of the transit midpoint (see Appendix~\ref{sec:appendix_Bayesian}).

\begin{figure}
\centering
\includegraphics[width=0.95\linewidth]{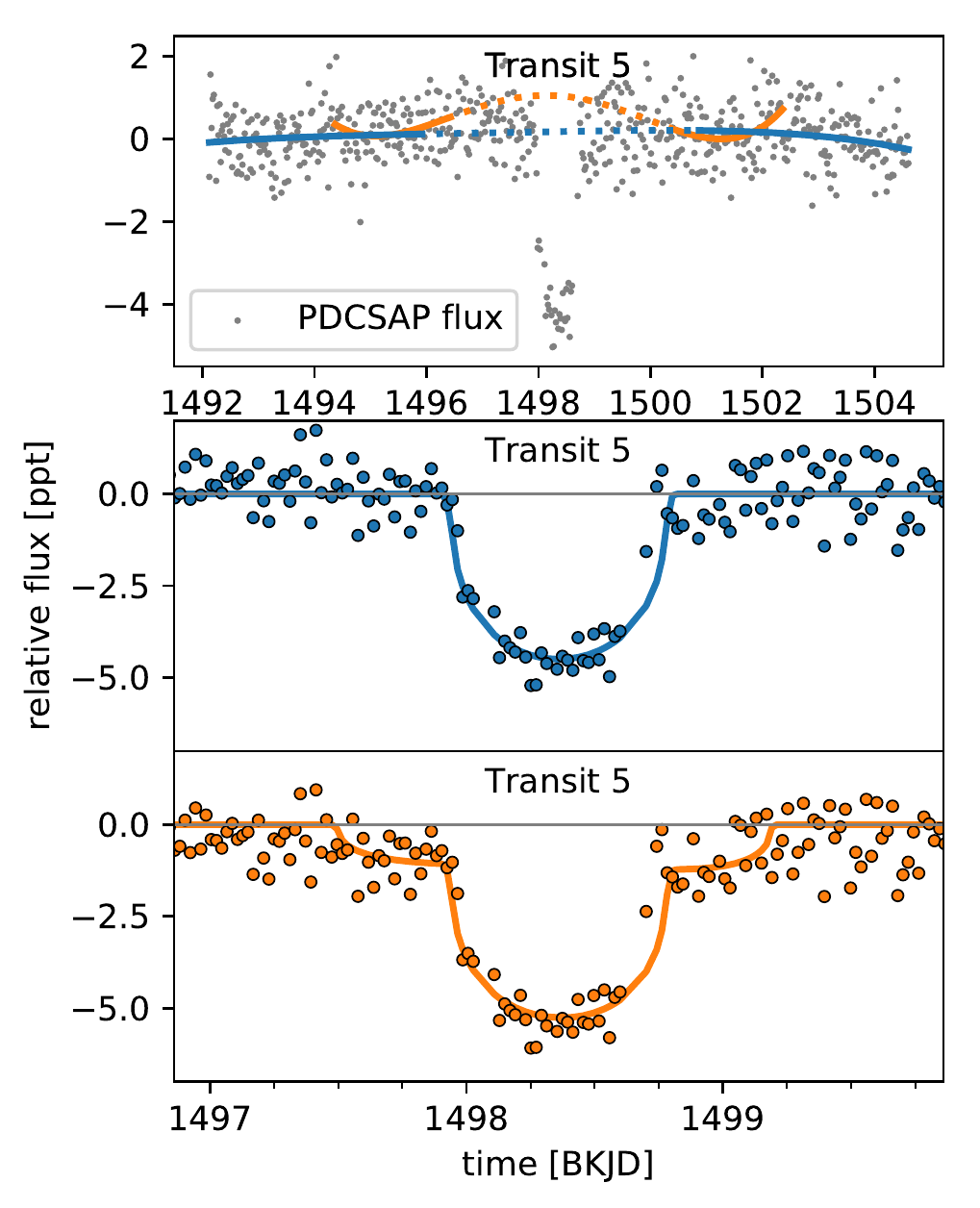}
\caption{Example of how the detrending procedure alone can produce an exomoon-like transit signal around a planetary transit. We use transit 5 of Kepler-1625\,b as an example. \textit{Top:} Gray dots indicate the Kepler PDCSAP flux. The lines show a 4th-order polynomial fit for which we exclude  $7.5$\,d (blue) or $4$\,d (orange) of data around the mid-point (dashed parts), respectively. \textit{Center:} Dots show the detrended light curve derived from the blue polynomial fit in the top panel. The blue line illustrates a planet-only transit model. \textit{Bottom:} Dots visualize the detrended light curve using the orange polynomial fit from the top panel. Note the additional moon-like transit feature caused by the overshooting of the orange polynomial in the top panel. The orange line shows a planet-moon transit model with moon parameters as in Table~\ref{tab:parameters_and_example} (see Fig.~\ref{fig:transit_slice} for transit dynamics). As an alternative interpretation, the blue detrending function filters out an actually existing moon signature while the orange detrending fit preserves the moon signal.}
\label{fig:detrending}
\end{figure}

Although this window length is astrophysically plausible to protect possible exomoon signals, many other choices are similarly plausible but they result in significantly different detrendings. Figure~\ref{fig:detrending} illustrates the effect on the detrended light curve if two different windows around the midpoint of the planetary transit (here transit 5) are excluded from the fitting. We chose a fourth-order polynomial detrending function and a $7.5$\,d (blue symbols) or a $4$\,d (orange symbols) region around the midpoint to be excluded from the detrending, mainly for illustrative purposes. In particular, with the latter choice we produce a moon-like signal around the planetary transit similar to the moon signal that appears in transit 5 in \cite{2018AJ....155...36T}. For the former choice, however, this signal does not appear in the detrended light curve.

\citet{2018AJ....155...36T} use the Cosine Filtering with Autocorrelation Minimization (CoFiAM) detrending algorithm to detrend both the SAP and PDCSAP flux around the three transits of Kepler-1625~b. CoFiAM fits a series of cosines to the light curve, excluding a specific region around the transit. CoFiAM preserves the signal of interest by using only cosines with a period longer than a given threshold and therefore avoids the injection of artificial signals with periods shorter than this threshold. \citet{2018AJ....155...36T} also test polynomial detrending functions but report that this removes the possible exomoon signal. We choose to reimplement the CoFiAM algorithm as our primary detrending algorithm to remain as close as possible in our analysis to the work in \citet{2018AJ....155...36T}. In our injection-retrieval test we also use polynomials of second, third, and fourth order for detrending. While low-order polynomials cannot generally fit the light curve as well as the series of cosines, the risk of injecting artificial signals may be reduced.

\subsubsection{Trigonometric detrending}
\label{sec:trig_detrending}
\begin{figure*}
\centering
\includegraphics[width=0.95\linewidth]{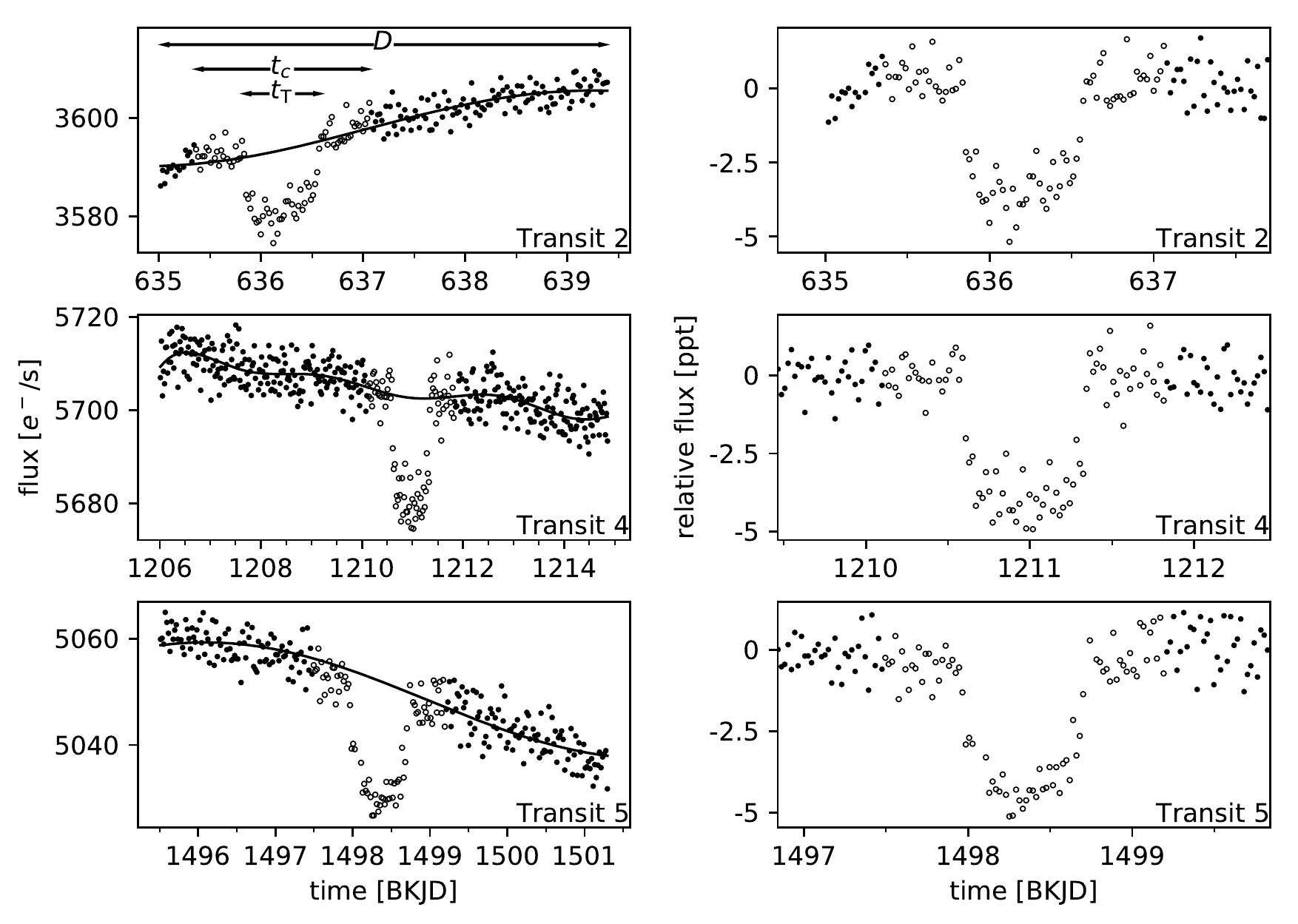}
\caption{\textit{Left:} Kepler SAP flux around the transits used for the trigonometric detrending, our reimplementation of the CoFiAM algorithm. The data points denoted by open circles around the transits are excluded from the detrending fit. The black line shows the resulting light curve trend without the transit. \textit{Right:} Detrended transit light curves as calculated by the trigonometric detrending.}
\label{fig:detrended_SAP}
\end{figure*}

We implement the CoFiAM detrending algorithm as per the descriptions given by \citet{2013ApJ...770..101K} and \cite{2018AJ....155...36T}. In the following, we refer to this reimplementation as trigonometric detrending as opposed to polynomial functions that we test as well (see Sect. \ref{sec:injection_detrending}).

The light curves around each transit are detrended independently.
For each transit, we start by using the entire SAP flux of the corresponding quarter. We use the SAP flux instead of the PDCSAP flux to reproduce the methodology of \citet{2018AJ....155...36T} as closely as possible. The authors argue that the use of SAP flux avoids the injection of additional signals into the light curve that might have the shape of a moon signal. First, we remove outliers using a running median with a window length of 20~h and a threshold of 3 times the local standard deviation with the same window length. In order to achieve a fast convergence of our detrending and transit fitting procedures, we initially estimate the transit midpoints and durations by eye and identify data anomalies, e.g. gaps and jumps (e.g. the jump 2\,d prior to transit 2 and the gap 4\,d after transit 4, see Fig.~\ref{fig:SAP_PDCSAP}).

Jumps in the light curve can have multiple reasons. The jumps highlighted around transit 2 in Fig.~\ref{fig:SAP_PDCSAP} are caused by a reaction wheel zero crossing event; the jump 5\,d after transit 4 is caused by a change in temperature after a break in the data collection. Following \citet{2018AJ....155...36T}, who ignore data points beyond gaps and other anomalous events for detrending, we cut the light curve around any of the transits as soon as it encounters the first anomaly, leaving us with a light curve of a total duration $D$ around each transit (see top left panel in Fig.~\ref{fig:detrended_SAP}). In Sect.~\ref{sec:injection_detrending}, we investigate the effect of including data beyond gaps. The detrending is then applied in two passes, using the first pass to get accurate transit parameters. In particular, we determine the duration ($t_{\rm T}$) between the start of the planetary transit ingress and the end of the transit egress \citep{2003ApJ...585.1038S} and the second pass to generate the detrended light curve.

\textit{First pass:} Using the estimated transit midpoints and durations, we calculate the time window ($t_{\rm c}$, see top left panel in Fig.~\ref{fig:detrended_SAP}) around a given transit midpoint to be cut from the detrending fit as $t_c=f_{t_{\rm c}}t_{\rm T}$, where the factor $f_{t_{\rm c}}$, relating the time cut around the transit to the transit duration, is an input parameter for the detrending algorithm. Specifically, $t_{\rm c}$ denotes the total length of time around the transit excluded from the detrending. We fit the detrending function
\begin{align}
\TrigFlux{k}(t,\vv{a},\vv{b})=a_0+\sum_{l=1}^k a_l\cos\left(l\frac{2\pi}{2 D}t\right)+b_l\sin\left(l\frac{2\pi}{2 D}t\right)
\end{align}
\noindent
 to the light curve (excluding the region $t_{\rm c}$ around the transit) by minimizing the $\chi^2$ between the light curve  and $\TrigFlux{k}(t,\vv{a},\vv{b})$, where $\vv{a}~=~(a_0,a_1, ..., a_k)$ and $\vv{b}~=~(b_1,b_2, ..., b_k)$ are the free model parameters to be fitted. The parameter $k$ is a number between 1 and $k_\text{max}=\text{round}(2D/t_{\rm p})$, where $t_{\rm p}=f_{t_{\rm p}}t_{\rm T}$ is the time scale below which we want to preserve possible signals. $f_{t_{\rm p}}$ is an input parameter to the detrending algorithm.
For each $k$ we divide the light curve by $\TrigFlux{k}$, giving us the detrended light curves $\DetrendedFlux^k$. We calculate the first-order autocorrelation according to the \cite{10.2307/2332391} test statistic for each $\DetrendedFlux^k$ (excluding again the region around the transit). For each transit we select the $\DetrendedFlux^k$ with the lowest autocorrelation $\DetrendedFlux^{k}_{\rm min}$ and combine these $\DetrendedFlux^{k}_{\rm min}$ around each transit into our detrended light curve $\DetrendedFlux$.
We fit the planet-only transit model to the detrended light curve $\DetrendedFlux$ and compute the updated transit midpoints and duration $t_{\rm T}$.

\textit{Second pass:} The second pass repeats the steps of the first pass, but using the updated transit midpoints and durations as input. The resulting detrended light curve $\DetrendedFlux$ is then used for our model fits with the ultimate goal of assessing whether an exomoon is a likely interpretation of the light curve signatures or not. We estimate the noise around each transit by taking the variance of $\DetrendedFlux$, excluding the transit region.

Figure \ref{fig:detrended_SAP} shows the detrending function as well as the detrended light curve for $f_{t_c}=2.2$ and $f_{t_p}=4.4$, corresponding to $t_c=1.6~$d and $t_p=3.1~$d.

\subsection{Transit model}
\label{sec:transit_model}

We construct two transit models, one of which contains a planet only and one of which contains a planet with one moon. We denote the planet-only model as $\mathcal{M}_0$ (the index referring to the number of moons) and the planet-moon model as $\mathcal{M}_1$. We do not consider models with more than one moon.

\subsubsection{Planet-only model}

$\mathcal{M}_0$ assumes a circular orbit of the planet around its star. Given the period of that orbit ($P$) and the ratio between stellar radius and the orbital semimajor axis ($R_\star/a$), the sky-projected apparent distance to the star center relative to the stellar radius can be calculated as

\begin{align}
z=\sqrt{\left[\frac{a}{R_\star}\sin\left(\frac{2\pi (t-t_0)}{P}\right)\right]^2+\left[b\cos\left(\frac{2\pi (t-t_0)}{P}\right)\right]^2} \ \ ,\label{eq:projected_distance}
\end{align}

\noindent
where $b$ is the transit impact parameter and $t_0$ is the time of the transit midpoint. We use the \texttt{python} implementation of the \cite{2002ApJ...580L.171M} analytic transit model by Ian Crossfield\footnote{Available at \href{http://www.astro.ucla.edu/\~ianc/files}{http://www.astro.ucla.edu/\textasciitilde{}ianc/files} as python.py.} to calculate the transit light curve based on the planet-to-star radius ratio ($r_{\rm p}=R_{\rm p}/R_\star$) and based on a quadratic limb-darkening law paramterized by the limb-darkening parameters $q_1$ and $q_2$ as given in \cite{2013MNRAS.435.2152K}. We call this model light curve with zero moons $\ModelFlux{0}$.

\begin{figure*}
\centering
\includegraphics[width=0.95\linewidth]{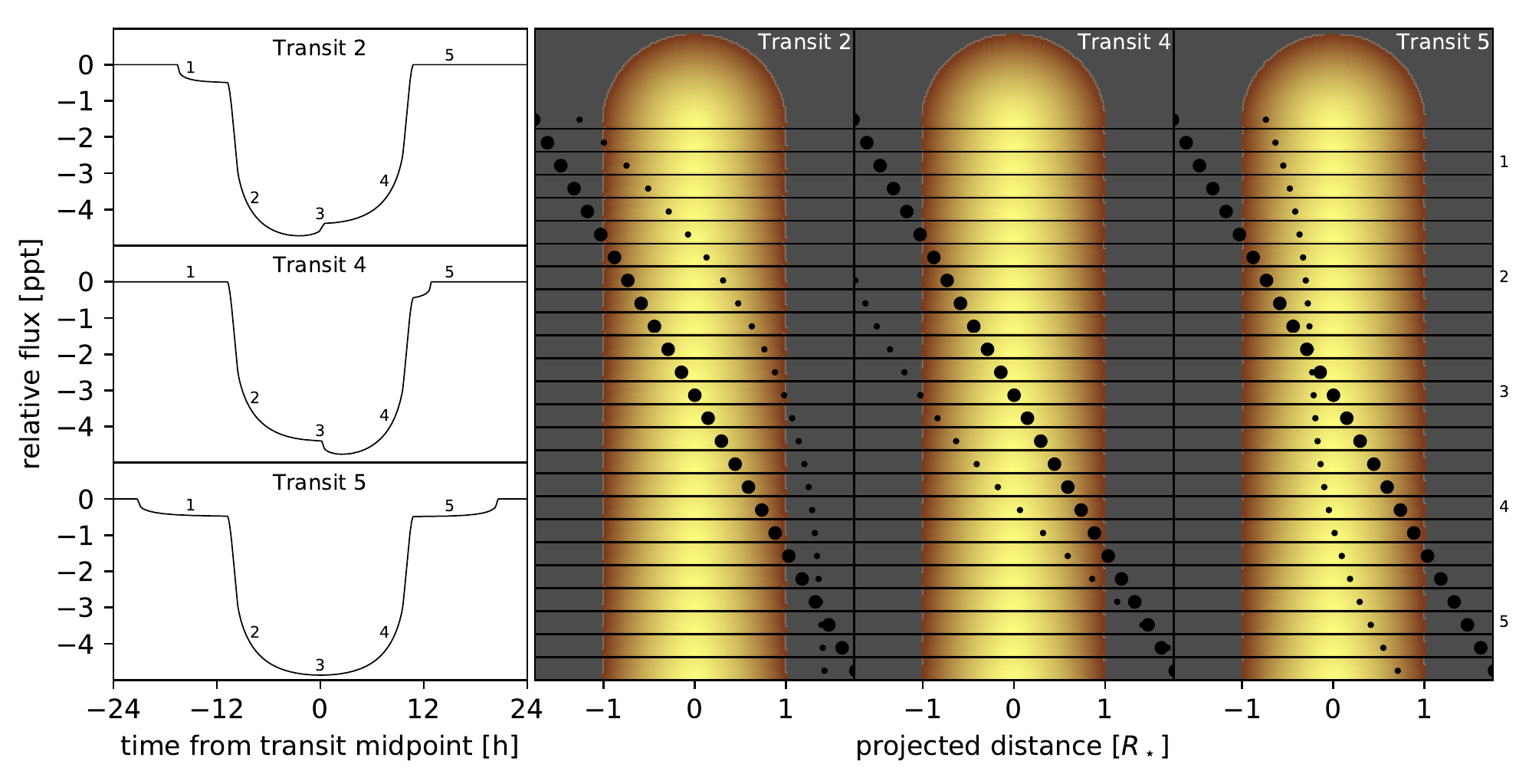}
\caption{\textit{Left}: Example of a simulated planet-moon transit light curve for transits 2, 4, and 5 using the nominal parameterization given in Table~\ref{tab:parameters_and_example}. The relative flux is the difference to the out-of-transit model flux and is given in parts per thousand (ppt). \textit{Right}: Visualization of the orbital configurations during transits 2 (left column), 4 (center column), and 5 (right column). Labels 1-5 in the light curves refer to configurations 1-5 (see labels along the vertical axis). An animation of this figure is available online.}
\label{fig:transit_slice}
\end{figure*}
\subsubsection{Planet-moon model}
\label{sec:onemoon}

\begin{table}
\center
\caption{Nominal parameterization of the planet-moon model to reproduce the transit shape suggested by \cite{2018AJ....155...36T}. The no-moon model uses the same parameter set (excluding the moon parameters), except that $R_\star$ and $a_{\rm B}$ are combined into a single parameter $R_\star/a_{\rm B}$.}
\def\arraystretch{1.2}
\begin{tabular}{c c p{0.45\columnwidth}}
\hline \hline
Parameter & Nominal Value & Description \\
\hline
$r_\text{p}$ & 0.06 & planet-to-star radius ratio\\
$a_{\rm B}$ & 0.9~au & circumstellar semimajor axis of the planet-moon barycenter\\
$b$ & 0.1 & planetary transit impact parameter\\
$t_{0,\rm B}$ & 61.51~d & transit midpoint of the planet-moon barycenter\\
$P_{\rm B}$ & 287.35~d& circumstellar orbital period of the planet-moon barycenter\\
\hline
$R_\star$ & 1.8~$R_\odot$ & stellar radius\\
$q_1$ & 0.6 & 1st limb darkening coefficient\\
$q_2$ & 0.2 & 2nd limb darkening coefficient\\
\hline
$r_\text{s}$ & 0.02 & moon-to-star radius ratio\\
$a_\text{s}$ & 1871~$R_{\rm J}$ & orbital semimajor axis of the planet-moon binary\\
$t_{0,\text{s}}$ & 1.86~d & time of planet-moon conjunction
\end{tabular}
\label{tab:parameters_and_example}
\end{table}

In our planet-moon model, we assume a circular orbit of the local planet-moon barycenter around the star with an orbital period $P_{\rm B}$, a semimajor axis $a_{\rm B}$, and a barycentric transit midpoint time $t_{0,\rm B}$. The projected distance of the barycenter to the star center relative to the stellar radius is calculated the same way as in eq. \ref{eq:projected_distance}. The planet and moon are assumed to be on circular orbits around their common center of mass with their relative distances to the barycenter determined the ratio of their masses $M_{\rm p}$ and $M_{\rm s}$ to the total mass $M_{\rm p}+M_{\rm s}$. The individual orbits of both the planet and the moon are defined by the total distance between the two objects $a_\text{ps}$, the planet mass $M_{\rm p}$, the moon mass $M_{\rm s}$ and by the time of the planet-moon conjunction $t_{0,\rm s}$, that is, the time at which the moon is directly in front of the planet as seen from an observer on Earth.

This model is degenerate in terms of the sense of orbital motion of the moon. In other words, a given planet-moon transit light curve can be produced by both a prograde and a retrograde moon \citep{2014ApJ...791L..26L,2014ApJ...796L...1H}. We restrict ourselves to prograde moons. The planet mass is set to a nominal 10 Jupiter masses, as suggested by \citet{2018AJ....155...36T} and in agreement with the estimates of \citet{2018A&A...610A..39H}. This constraint simplifies the interpretation of the results substantially since the moon parameters are then unaffected by the planetary parameters. The moon mass is assumed to be much smaller than that of the planet. In fact, for a roughly Neptune-mass moon around a 10 Jupiter-mass planet, we expect a TTV amplitude of 3 to 4 minutes and a TDV amplitude of 6 to 7 minutes, roughly speaking. Hence, we simplify our model and set $M_{\rm s}=0$, which means that $a_{\rm ps}$ is equal to the distance between the moon and the planet-moon barycenter, $a_{\rm s}$. The moon is assumed to have a coplanar orbit around the planet and, thus, to have the same transit impact parameter as the planet.

With these assumptions the projected distance of the planet center to the star center relative to the stellar radius $z_{\rm p}$ is equal to that of the barycenter $z_{\rm B}$. The projected distance of the moon center to the star center relative to the stellar radius $z_{\rm s}$ is given by
\begin{align}
z_{\rm s}^2=&\left[\frac{a_{\rm B}}{R_\star}\sin\left(\frac{2\pi (t-t_{0,\rm B})}{P_{\rm B}}\right)+\frac{a_{\rm ps}}{R_\star}\sin\left(\frac{2\pi (t-t_{0,\rm s})}{P_{\rm s}}\right)\right]^2\nonumber\\
+&\left[b\cos\left(\frac{2\pi (t-t_{0,\rm B})}{P_{\rm B}}\right)\right]^2\ \ ,
\end{align}
where $P_{\rm s}$ is the orbital period of the moon calculated from the fixed masses and $a_{\rm ps}$.

We calculate the transit light curves of both bodies and combine them into the total model light curve, which we call $\ModelFlux{1}$. We use the limb-darkening parameter transformation from \cite{2013MNRAS.435.2152K}. For computational efficiency, we do not consider planet-moon occultations. For the planet-moon system of interest,  occultations would only occur only about half of the transits (assuming a random moon phase) even if the moon orbital plane would be perfectly parallel to the line of sight. Such an occultation would take about 1.5\,h and would only affect 5-10\,\% of the total moon signal duration.

In Table \ref{tab:parameters_and_example} we give an overview of our nominal parameterization of the planet-moon model. In Fig. \ref{fig:transit_slice} we show the orbital dynamics of the planet and moon during transits 2, 4, and 5 using the nominal parameters in Table~\ref{tab:parameters_and_example}. This nominal parameterization was chosen to generate a model light curve that is reasonably close to the preferred model light curve found in \citet{2018AJ....155...36T}, but it does not represent our most likely model fit to the data.

\subsubsection{Finding the posterior probability distribution}
\label{sec:posterior}

We use the Markov Chain Monte Carlo (MCMC) implementation {\tt Emcee} \citep{2013PASP..125..306F} to estimate the posterior probability distribution of the parameters for model $\Model{(i)}$ ($\NoMoonModel$ or $\OneMoonModel$). For this purpose, we need to formulate the probability density of a light curve as well as the prior of the parameters.

All three transits taken together, we have a total of $N$ detrended flux measurements (see Sect.~\ref{sec:trig_detrending}). Given a set of parameters $\ParamVec$, model $\Model{i}$ produces a model light curve $\ModelFlux{i}(t,\ParamVec)$. We assume that the noise is uncorrelated (see Appendix~\ref{sec:appendix_autocorrelation}) and Gaussian with a standard deviation $\sigma_j$ at time $t_j$. This simplifies the joint probability density to a product of the individual probabilities. The joint probability density function of the detrended flux $\DetrendedFlux(t)$ is given by
\begin{align}
p(\DetrendedFlux|\ParamVec,\Model{i})=\prod_{j=1}^{N} \frac{1}{\sqrt{2\pi\sigma_j^2}}\exp\left(-\frac{\left(\DetrendedFlux(t_j)-\ModelFlux{i}(t_j,\ParamVec)\right)^2}{2\sigma_j^2}\right)\;.
\end{align}

\noindent
The noise dispersion $\sigma_j$ has a fixed value for each transit.

Table~\ref{tab:param_range} shows the parameter ranges that we explore. A prior is placed on the stellar mass according to the mass of $1.079^{+0.100}_{-0.138}\;M_\odot$ determined by \cite{2017ApJS..229...30M}. The stellar mass for a given parameter set is determined from the system's total mass using $P_{\rm B}$ and $a_{\rm B}$ and subtracting the fixed planet mass of 10 Jupiter masses.

A total of 100 walkers are initiated with randomly chosen parameters close to the estimated transit parameters. For the sake of fast computational performance, the walkers are initially separated into groups of 16 for the planet-only model and 24 for the planet-moon model (twice the number of parameters plus 2, respectively), temporarily adding walkers to fill the last group. To transform the initially flat distribution of walkers into a distribution according to the likelihood function, the walkers have to go through a so-called burn-in phase, the resulting model fits of which are discarded. We chose a burn-in phase for the walkers of 500 steps in both groups. Afterwards, we discard the temporarily added walkers, merge the walkers back together, and perform a second burn-in phase of 2\,200 steps with a length determined by visual inspection. Finally, we initiate the main MCMC run with a total of 8\,000 steps.

We run the MCMC code on the detrended light curve using both the planet-only and the planet-moon model.

\begin{table}
\center
\caption{Parameter ranges explored with our planet-moon model. The ranges of the no-moon model parameters are the same for the shared parameter and is propagated to the derived parameter $R_\star/a$.}
\def\arraystretch{1.2}
\begin{tabular}{c c c c c}
\hline \hline
Min. Value & & Parameter & & Max. Value \\
\hline
0              & $\leq$ & $r_{\rm p}$ & $\leq$ & 0.1            \\
0.2\,au        & $\leq$ & $a_{\rm B}$ & $\leq$ & 2\,au          \\
0              & $\leq$ & $b$         & $\leq$ & 1              \\
$-P_{\rm B}/2$ & $\leq$ & $t_{0,\rm B}$   & $\leq$ & $P_{\rm B}/2$ \\
270\,d         & $\leq$ & $P_{\rm B}$     & $\leq$ & 300\,d         \\
\hline
0              & $\leq$ & $R_\star$   & $\leq$ & $4.3\,R_\odot$ \\
0              & $\leq$ & $q_1$       & $\leq$ & 1              \\
0              & $\leq$ & $q_1$       & $\leq$ & 1              \\
\hline
0              & $\leq$ & $r_{\rm s}$ & $\leq$ & $r_{\rm p}$    \\
0              & $\leq$ & $a_{\rm s}$ & $\leq$ & $R_{\rm Hill} / 2$\\
$-P_{\rm s}/2$ & $\leq$ & $t_{0,\rm s}$   & $\leq$ & $P_\text{s}/2$
\end{tabular}
\label{tab:param_range}
\end{table}

\subsection{Model selection}

We use the Bayesian Information Criterion (BIC) to evaluate how well a model describes the observations in relation to the number of model parameters and data points. The BIC of a given model $\Model{i}$ with $m_{i}$ parameters is defined by \citet{1978AnSta...6..461S} as

\begin{align}
 \text{BIC}(\Model{i}|\DetrendedFlux)=m_{i}\ln N-2\ln p(\DetrendedFlux|\ParamVec_\text{max},\Model{i}),
\end{align}

\noindent
where $\ParamVec_\text{max}$ is the set of parameters that maximizes the probability density function $p(\DetrendedFlux|\ParamVec,\Model{i})$ for a given the light curve $\DetrendedFlux$ and model $\Model{i}$.

The difference of the BICs between two models gives an indication as to which model is more likely. In particular, $\Delta\text{BIC}(\OneMoonModel,\NoMoonModel)~\equiv~\text{BIC}(\OneMoonModel)~-~\text{BIC}(\NoMoonModel)~<~0$ if model $\OneMoonModel$ is more likely. We consider $\Delta{\rm BIC}~<~6$ (or $\Delta{\rm BIC}~>~6$) as strong evidence in favor of (or against) model $\OneMoonModel$ \citep[see, e.g.,][]{kass_1995}.

The best-fitting set of parameters derived from our MCMC runs ($\vv{\theta}_{\text{max}}$) is then used to calculate $\Delta\text{BIC}(\OneMoonModel,\NoMoonModel)$. For our calculations, we only use those parts of the light curve around the transits that could potentially be affected by a moon ($3.25~$d on each side of the transits, determined by the Hill radius $R_\text{Hill}$ and the orbital velocity of the planet-moon barycenter, see Appendix~\ref{sec:appendix_Bayesian}).

\subsection{Injection-retrieval test}
\label{sec:injection_retrieval}

In order to estimate the likelihood of an exomoon feature to be due to either a real moon or due to noise, we perform several injection-retrieval experiments. One of us (MH) injected two cases of transits into the out-of-transit parts of the original PDCSAP Kepler flux. In one case, a sequence of three planet-only transits (similar to the sequence of real transits 2, 4, and 5) was injected, where the planet was chosen to have a radius of 11 Earth radii. In another case, a sequence of three transits of a planet with moon with properties similar to the proposed Jupiter-Neptune system was injected. Author KR then applied the Baysian framework described above in order to evaluate the planet-only vs. the planet-with-moon hypotheses and in order to characterize the planet and (if present) its moon.

As an important trait of our experiment, KR did not know which of the light curves contained only a planet and which contained also a moon.

\subsubsection{Transit injections into light curves}
\label{sec:model_injection}

For the injection part, we use {\tt PyOSE} \citep{2016ApJ...820...88H,2016A&A...591A..67H} to create synthetic planet and moon ensemble transits. This code numerically integrates the non-occulted areas of the stellar disk to calculate the instantaneous flux of the star, which makes it a computationally slow procedure. Hence we use the analytical model described Sect. \ref{sec:transit_model} for the retrieval part. In our model, the moon's orbit is defined by its eccentricity ($e_{\rm s}$, fixed at 0), $a_{\rm s}$, its orbital inclination with respect to the circumstellar orbit ($i_{\rm s}$, fixed at $90^\circ$), the longitude of the ascending node, the argument of the periapsis, and the planetary impact parameter ($b$, fixed at zero). Due to the small TTV and TDV amplitudes compared to the 29.4\,min exposure of the Kepler long cadence data, we neglect the planet's motion around the planet/moon barycenter, although {\tt PyOSE} can model this dynamical effect as well, and assume that the moon orbits the center of the planet.

Our numerical code creates a spherical limb-darkened star on a 2-dimensional grid of floating-point values. The sky-projected shapes of both the planet and the moon are modeled as black circles. The spatial resolution of the simulation is chosen to be a few million pixels so that the resulting light curve has a numerical error $<1$\,ppm that is negligible compared to the $\approx\,700$\,ppm noise level of the Kepler light curve. The initial temporal resolution of our model is equivalent to 1\,000 steps per planetary transit duration, which we then downsample to the observed 29.4\,min cadence. The creation of one such light curve of a planet with a moon takes about one minute on a modern desktop computer.

We create a set of 100 such transit simulations of the planet-moon ensemble, where the two bodies move consistently during and between transits. All orbits are modeled to be co-planar and mutual planet-moon occultations are also included. For each transit sequence, the initial orbital phase of the planet-moon system is chosen randomly.

With $P_{\rm B}=287.378949$\,d and $P_{\rm s}=2.20833$\,d, the moon advances by $\approx0.13$ in phase between each subsequent planetary transit ($P_{\rm B}/P_{\rm s} \approx 130.13$). During a planetary transit, the moon advances by $\approx0.36~$rad in phase (the planetary transit duration is $0.7869\pm0.0084$\,d).

We also create a set of 100 such transits that only have a transiting planet without a moon. In these cases, the planetary radius was increased slightly to match the average transit depth of planet and moon.

\subsubsection{Testing the model-selection algorithm on synthetic light curves with white noise only}

\begin{figure*}
\centering
\includegraphics[width=0.95\linewidth]{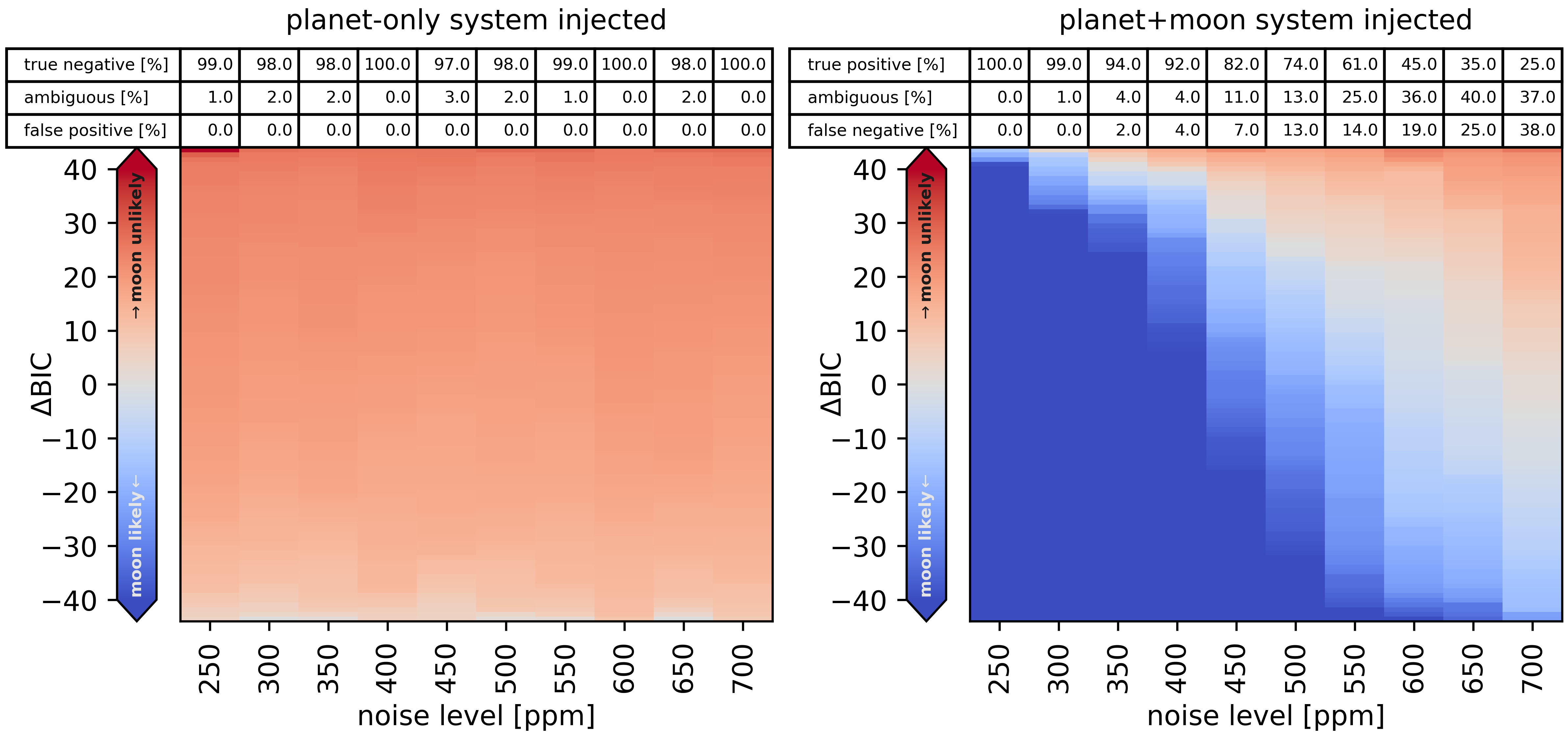}
\caption{Difference between the BIC of the planet-moon model and the no-moon model for 2\,000 artificial white noise light curves at different noise levels, injected with simulated transits. On the left ($100\times10$ light curves) a planet and moon transit was injected, on the right ($100\times10$ light curves) only the planet. Each light curve consists of three consecutive transits. Each column is sorted by the $\Delta$BIC. The $\Delta$BIC threshold, over which a planet-moon or planet-only system is clearly preferred is $\pm6$ with the state of systems with a $\Delta$BIC between those values considered to be ambiguous.}
\label{fig:white_noise_delta_bic}
\end{figure*}
\begin{figure*}
\centering
\includegraphics[width=0.95\linewidth]{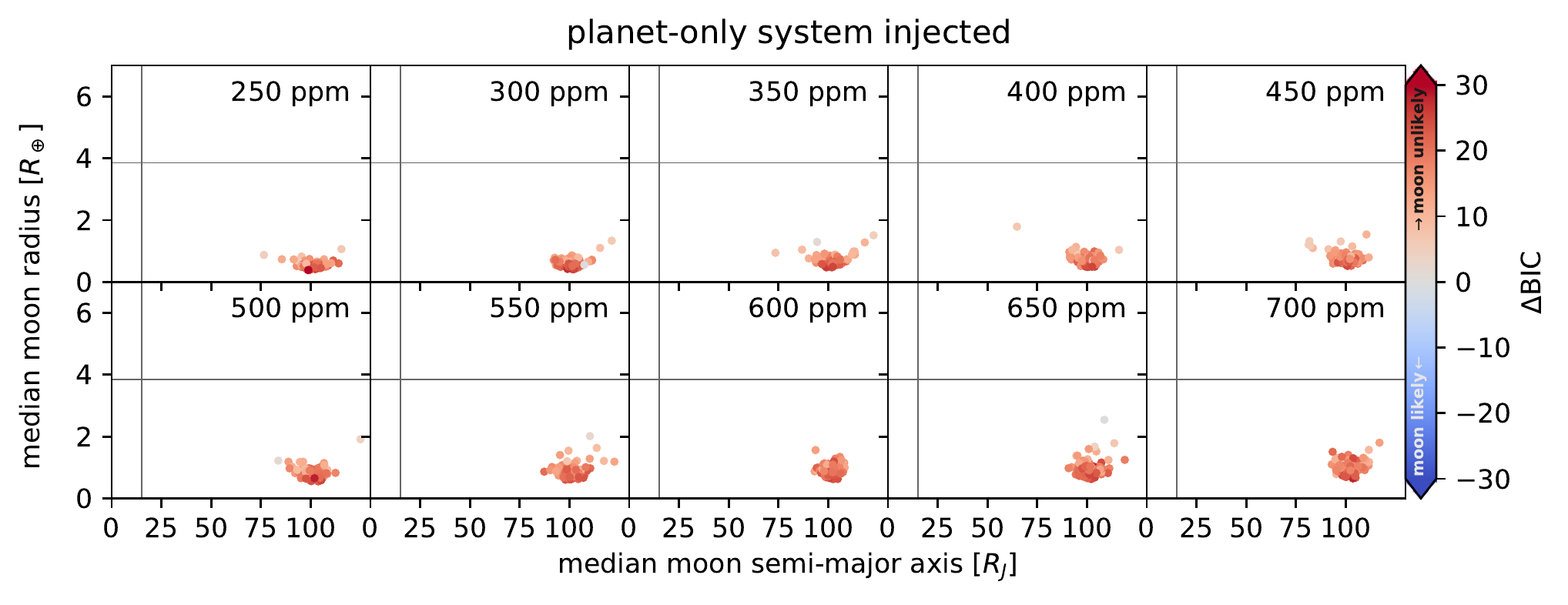}
\includegraphics[width=0.95\linewidth]{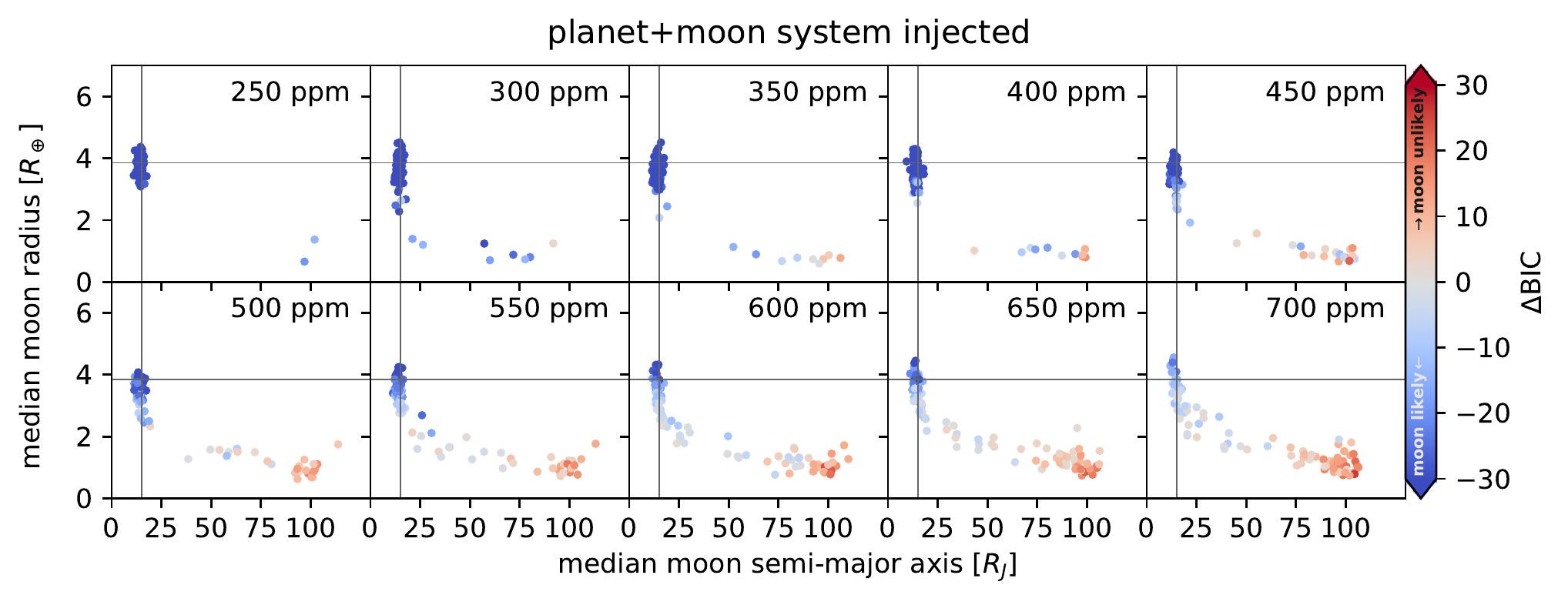}
\caption{Distribution of the median likelihood $R_{\rm s}$ and $a_{\rm s}$ for all the runs for the different noise levels, with the runs injecting planet and moon on the top and runs injecting only a planet in the bottom. The $\Delta$BIC of the planet-moon model compared to the no-moon model for all runs is indicated by the color. Generally runs with a low $\Delta$BIC (indicating the presence of a moon) also are in the vicinity of the injected parameter.}
\label{fig:R_M_a_M_evolution_with_noise_level}
\end{figure*}
As a first validation of our injection-retrieval experiment and our implementation of the Bayesian statistical framework, we generate a new set of white noise light curves to test only the model comparison part of our pipeline without any effects that could possibly arise from imperfect detrending. Any effects that we would see in our experiments with the real Kepler-1625 light curve but not in the synthetic light curves with noise only could then be attributed to the imperfect detrending of the time-correlated (red) noise.

MH generated 200 synthetic light curves with ten different levels of white noise, respectively, ranging from root mean squares of 250\;ppm to 700\;ppm in steps of 50\;ppm. This results in a total of 2\,000 synthetic light curves. MH used the method described in Sect.~\ref{sec:model_injection} to inject three transits of a planet only into 100 light curves per noise level and three transits of a planet with a moon into the remaining 100 light curves per noise level. The initial orbital phases were randomly chosen and are different from the ones used to generate the light curves in Sect.~\ref{sec:injection_real_light_curve}. MH delivered these light curves to KR without revealing their specific contents. KR then ran our model selection algorithm to find the $\Delta$BIC for each of the 2\,000 systems. After the $\Delta$BICs were found, MH revealed the planet-only or the planet-moon nature of each light curve.

Fig. \ref{fig:white_noise_delta_bic} shows the resulting $\Delta$BICs for each of the 2\,000 light curves, separated into the planet-only (left panel) and planet-moon injected systems (right panel) and sorted by the respective white noise level (along the abscissa). Each vertical column contains 100 light curves, respectively. For a noise level of 250\;ppm, as an example, our algorithm finds no false positive moons in the planet-only data, that is, no system with a ${\Delta}{\rm BIC}<-6$, while 1 case remains ambiguous ($-6~<~\Delta{\rm BIC}~<~6$) and the other 99 cases are correctly identified as containing no moons. In the case of an injected planet-moon system instead, the algorithm correctly retrieves the moon in 100\,\% of the synthetic light curves, that is, ${\Delta}{\rm BIC}<-6$ for all systems.

More generally, for the simulated planet-only systems, the false positive rate is 0\,\% throughout all noise levels. Occasionally a system is flagged as ambiguous, but overall the algorithm consistently classifies planet-only systems correctly as having no moon. Referring to the injected planet-moon system (right panel), our false negative rate rises steadily with increasing noise level. In fact, it reaches parity with the true positive rate between about 650\;ppm and 700\;ppm. 

In Fig.~\ref{fig:R_M_a_M_evolution_with_noise_level} we present $a_{\rm s}$ and $R_{\rm s}$ for each of the maximum-likelihood fits shown in Fig.~\ref{fig:white_noise_delta_bic}. Each panel in Fig.~\ref{fig:R_M_a_M_evolution_with_noise_level} refers to one white noise level, that is, to one column in Fig.~\ref{fig:white_noise_delta_bic} of either the planet-only or the planet-moon injected system. In the case of an injected planet only (upper panels), the most likely values of $a_{\rm s}$ are distributed almost randomly over the range of values that we explored. On the other hand, $R_{\rm s}$ is constrained to a small range from about $1.5\,R_\oplus$ at 250~ppm to roughly $3\,R_\oplus$ at 700~ppm with the standard variation naturally increasing with the noise level.

The lower part of Fig.~\ref{fig:R_M_a_M_evolution_with_noise_level} shows the outcome of our planet-moon injection-retrievals from the synthetic light curves with white noise only. The correct parameters are generally recovered at all noise levels. In fact, we either recover the moon with a similar radius and orbital separation as the injection values (symbolized by blue points) or we find the moon to have very different radius and orbit while also rejection the hypothesis of its presence in the first place (symbolized by red points). The distribution of these false negatives in the $a_{\rm s}$-$R_{\rm s}$ plane resembles the distribution of the true negatives in the corresponding no-moon cases. The ambiguous runs with a $\Delta$BIC around 0 still mostly recover the injected moon parameters. This is especially clear for the 700~ppm level, with 50\,\% more ambiguous runs than true positives, where most of the runs still recover the injected parameters.

\subsubsection{Transit injection into real out-of-transit data}
\label{sec:injection_real_light_curve}

We inject synthetic transits into the Kepler-1625 PDCSAP data prior to our own detrending (see Sect. \ref{sec:model_injection}). We use the PDCSAP flux instead of SAP flux because (1) it was easier for us to automate the anomaly detection and (2) PDCSAP flux has been cleaned from common systematics. Since the PDC pipeline removes many of the jumps in the data, we can focus on a single type of anomaly, that is gaps. Gaps are relatively easy to detect in an automated way, removing the requirement of visual inspection of each light curve. For the injection, we select out-of-transit parts of the Kepler-1625 light curve that have at least 50\,d of mostly uninterrupted data (25\,d to both sides of the designated time of transit injection), but accept the presence of occasional gaps with durations of up to several days during the injection process.

The set of 200 synthetic light curves was provided by MH to KR for blind retrieval without any disclosure as to which of the sequences have a moon. The time of mid-transit was communicated with a precision of 0.1\,days to avoid the requirement of a pre-stage transit search. This is justified because (i.) the original transits of Kepler-1625\,b have already been detected and (ii.) the transit are visible by-eye and do not necessarily need computer-based searches. We provide the 200 datasets to the community for reproducibility\footnote{Available on Zenodo, \href{http://doi.org/10.5281/zenodo.1202034}{[10.5281/zenodo.1202034]}, \citet{hippke_zenodo}} and encourage further blind retrievals.

\subsubsection{Detrending of the transit-injected light curves}
\label{sec:injection_detrending}

The detrending procedure for our injection-retrieval experiment differs from the one used to detrend the original light curve around the Kepler-1625\,b transits (see Sect.~\ref{sec:trig_detrending}) in two regards.

First, we test the effect of the detrending function. In addition to the trigonometric function, we detrend the light curve by polynomials of second, third and fourth order.

In addition, we test if the inclusion or neglect of data beyond any gaps in the light curve affects the detrending. In one variation of our detrending procedure, we use the entire $\pm~25$\,d of data (excluding any data within $t_{\rm c}$) around a transit midpoint. In another variation, we restrict the detrending to the data up to the nearest gap (if present) on both sides of the transit.

To avoid the requirement of time-consuming visual inspections of each light curve, we construct an automatic rule to determine the presence of gaps, which are the most disruptive kind of artifact to our detrending procedure. We define a gap as an interruption of the data of more than half a day. Whenever we do detect a gap, we cut another 12\,h at both the beginning and the end of the gap, since our visual inspection of the data showed that many gaps are preceded or followed by anomalous trends (see e.g. the gap 4\,d after transit 4 in Fig.~\ref{fig:SAP_PDCSAP}).

We ignore any data points within $t_{\rm c}$ around the transit midpoint (see Sect.~\ref{sec:trig_detrending}). If a gap starts within an interval $[t_{\rm c}/2,t_{\rm c}/2~+~12\,{\rm h}]$ around the transit midpoint, then we lift our constraint of dismissing a 12\,h interval around gaps and use all the data within $[t_{\rm c}/2,t_{\rm c}/2~+~12\,{\rm h}]$ plus any data up to 12\,h around the next gap.

If all these cuts result in no data points for the detrending procedure to one side of one of the three transits in a sequence, then we ignore the entire sequence for our injection-retrieval experiment. This is the case for \text{40} out of the 200 artificially injected light curves. This high loss rate of our experimental data is a natural outcome of the gap distribution in the original Kepler-1625 light curve. We exclude these 40 light curves for all variations of the detrending procedure that we investigate. All things combined, these constraints produce synthetic light curves with gap characteristics similar to the original Kepler-1625\,b transits (see Fig.~\ref{fig:detrended_SAP}), that is, we allow the simulation of light curves with gaps close to but not ranging into the transits. The four detrending functions and our two ways of treating gaps yield a total of eight different detrending methods that we investigate (see Table~\ref{tab:combination_detrend_methods_injected_data}).

\begin{table}
\center
\caption{Definition of the detrending identifiers in relation to the respective detrending functions that we explored in our transit injection-retrieval experiment of the Kepler-1625 data. We define a gap as any empty parts in the light curve that show more than 12\,h between consecutive data points. The trigonometric function refers to our reimplementation of the CoFiAM algorithm. \texttt{P2} to \texttt{P4} refer to polynomials of second to fourth order. \texttt{T} refers to our trigonometric detrending. \texttt{G} stands for the inclusion of data beyond gaps, \texttt{N} stands for the exclusion of data beyond gaps.}
\def\arraystretch{1.2}
\begin{tabular}{c c c}
\hline \hline
 Identifier & Detrending Function & Reject Data Beyond Gap? \\
 \hline
 \texttt{P2/G} & 2nd order polynomial & yes \\
 \texttt{P2/N} & 2nd order polynomial & no \\
 \texttt{P3/G} & 3rd order polynomial & yes \\
 \texttt{P3/N} & 3rd order polynomial & no \\
 \texttt{P4/G} & 4th order polynomial & yes \\
 \texttt{P4/N} & 4th order polynomial & no \\
 \texttt{T/G} & trigonometric & yes \\
 \texttt{T/N} & trigonometric & no \\
\end{tabular}
\label{tab:combination_detrend_methods_injected_data}
\end{table}

\section{Results}
\begin{figure}
\centering
\includegraphics[width=0.95\linewidth]{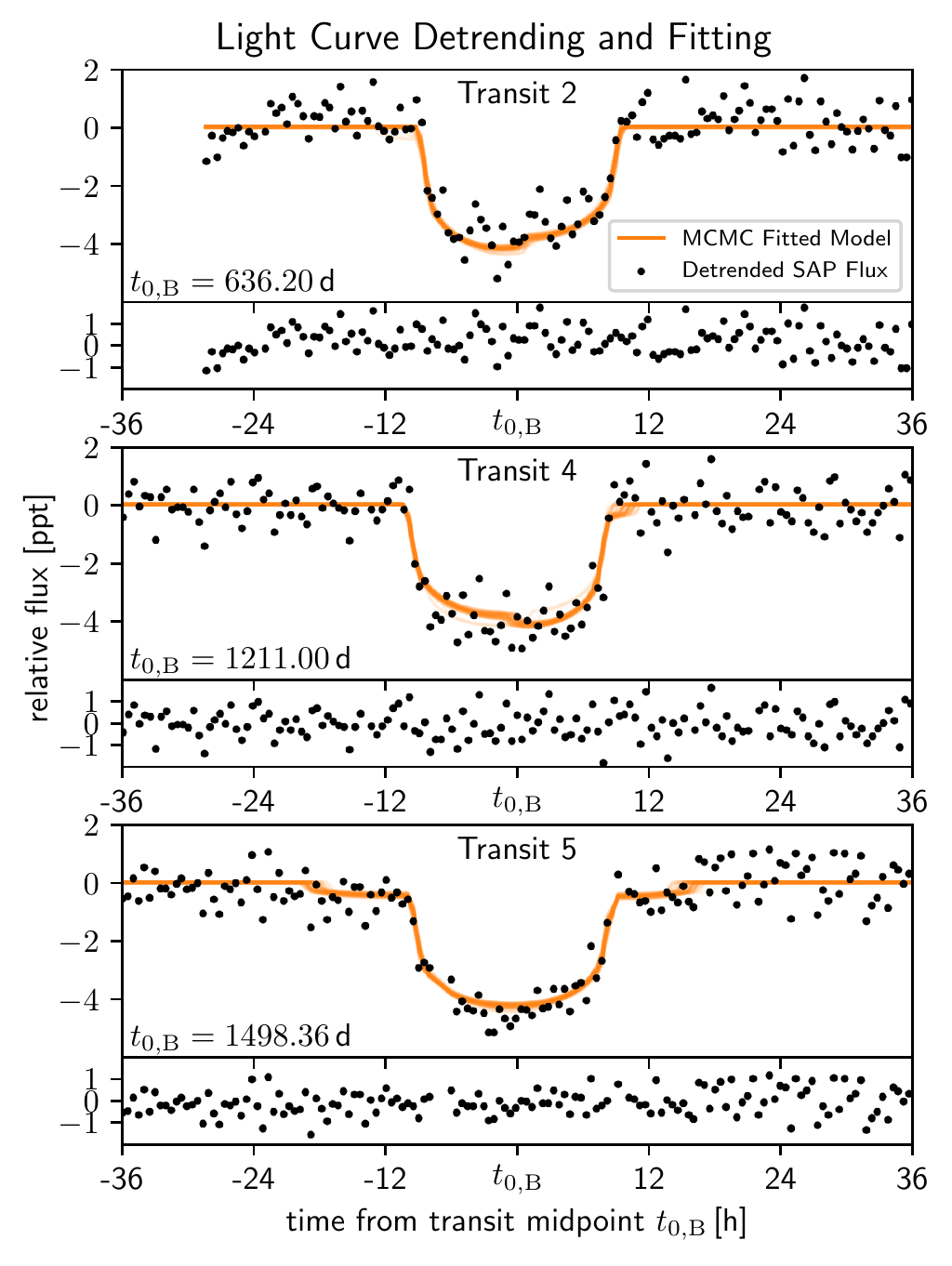}
\caption{The observed 2nd, 4th, and 5th transits of Kepler~1625\,b. Black dots refer to our detrended light curve from the trigonometric detrending procedure, and orange curves are the model light curves generated using the 100 best fitting parameter sets of the MCMC run. The $\Delta{\rm BIC}$, calculated from the most likely parameters, is $-4.954$.}
\label{fig:teachey_best_fits}
\end{figure}
\begin{figure}
\centering
\includegraphics[width=0.95\linewidth]{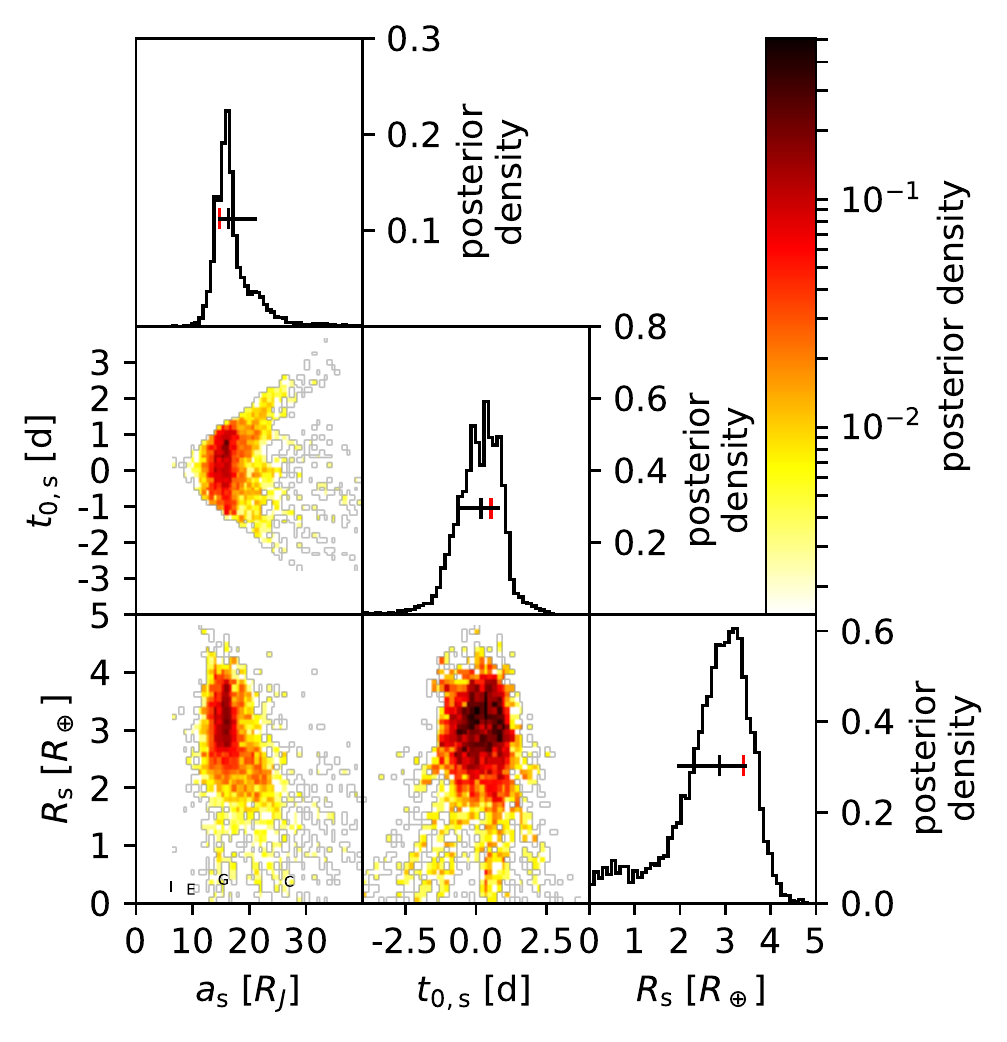}
\caption{Posterior probability distribution of the moon parameters generated by the MCMC algorithm for the light curve detrended by the trigonometric detrending. The black vertical lines show the median of the posterior distribution, the black horizontal lines indicate the $1\sigma$ range around the median. The red vertical lines show the point of maximum likelihood. The locations of the Galilean moons are included in the lower left panel for comparison.}
\label{fig:teachey_small_grid}
\end{figure}

Our first result is a reproduction of a detrended transit light curve of Kepler-1625\,b that has the same morphology and moon characterization as the one proposed by \cite{2018AJ....155...36T} and that has a negative $\Delta$BIC. We explore the variation of the free parameters of our trigonometric detrending procedure, $f_{t_p}$ and $f_{t_c}$, and identify such a detrended light curve for $f_{t_p}=4.4$ and $f_{t_c}=2.2$. Figure~\ref{fig:teachey_best_fits} shows the resulting light curve.

In Fig.~\ref{fig:teachey_small_grid} we show the results of our MCMC analysis of this particular light curve, which yields a moon with $a_\text{s}=16.3^{+5.0}_{-1.9}~R_{\rm J}$ and $R_\text{s}=2.87^{+0.61}_{-0.94}~R_\oplus$. While both the moon radius and semimajor axis are well constrained, the distribution of the initial planet-moon orbital conjunction ($t_{0,{\rm s}}$) fills out almost the entire allowed parameter range from $-1/2~P_{\rm s}$ to $+1/2~P_{\rm s}$. The planetary radius is $0.863^{+0.072}_{-0.051}~R_{\rm J}$, the stellar radius is $R_\star=1.57^{+0.11}_{-0.09}~R_\odot$, and the density is $\rho_\star=0.26^{+0.04}_{-0.05}~\rho_{\odot}$.

The point of maximum likelihood in the resulting MCMC distribution is at $a_\text{s}=14.7~R_{\rm J}$, $R_\text{s}=3.4~R_\oplus$, $R_\star=1.57~R_\odot$, $\rho_\star=0.23~\rho_{\odot}$ and $R_{\rm p}=8.63~R_{\rm J}$. The $\Delta \text{BIC}(\mathcal{M}_1,\mathcal{M}_0)$ we found is -4.954, indicating moderate evidence in favor of an exomoon being in the light curve.

\subsection{Injection-retrieval experiment}

\begin{figure*}[h!]
\centering
\includegraphics[width=0.95\linewidth]{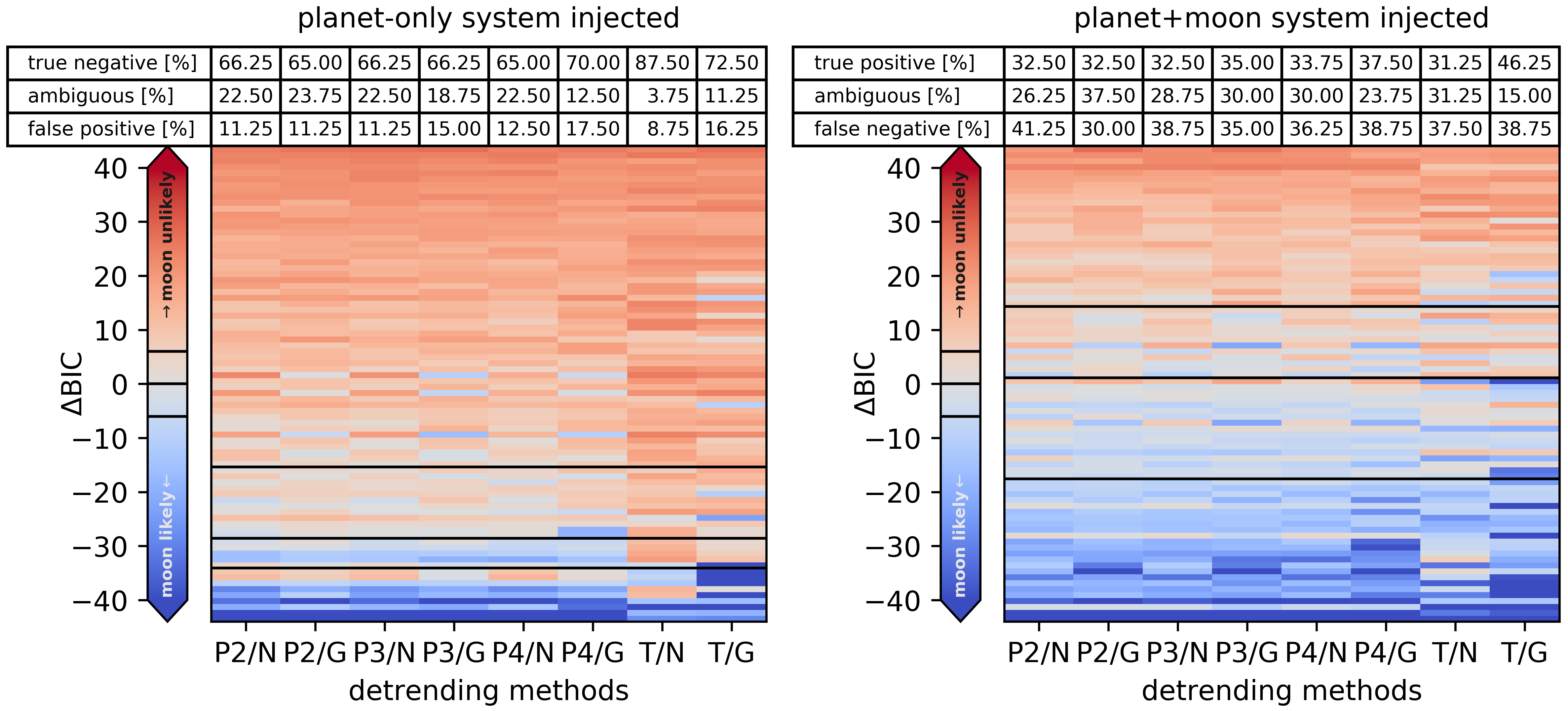}
\caption{Difference between the BIC of the planet-moon model and the no-moon model using different detrending methods for 160 light curves, generated using the PDCSAP flux of Kepler~1625, injected with three simulated transits. On the left (80 light curves) a planet and moon transit was injected, on the right (80 light curves) only the planet. Each light curve consists of three consecutive transits. Each row of 8 detrending methods uses the same light curve. The rows are sorted by their mean $\Delta$BIC, with black lines indicating the $\Delta{\rm BIC}=\{-6, 0, 6\}$ positions for the mean $\Delta{\rm BIC}$ per row.}
\label{fig:injection_delta_bic}
\end{figure*}
\begin{figure*}
\centering
\includegraphics[width=0.95\linewidth]{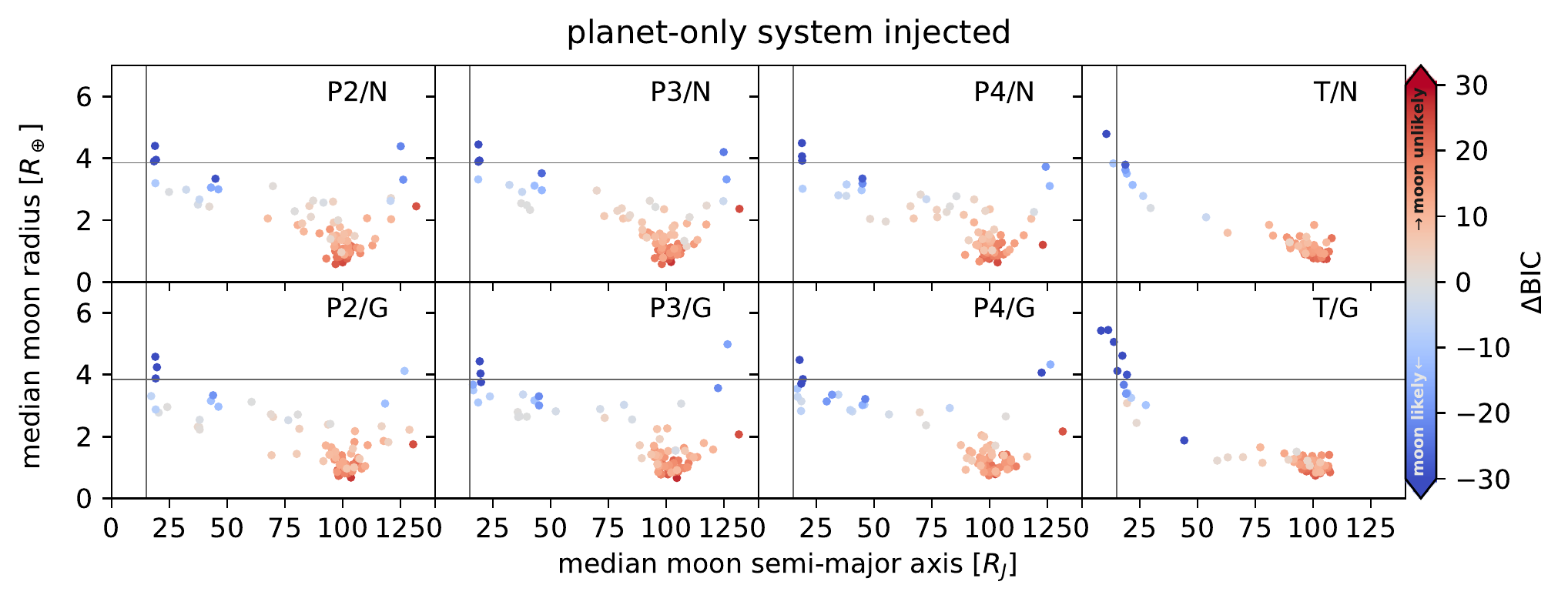}
\includegraphics[width=0.95\linewidth]{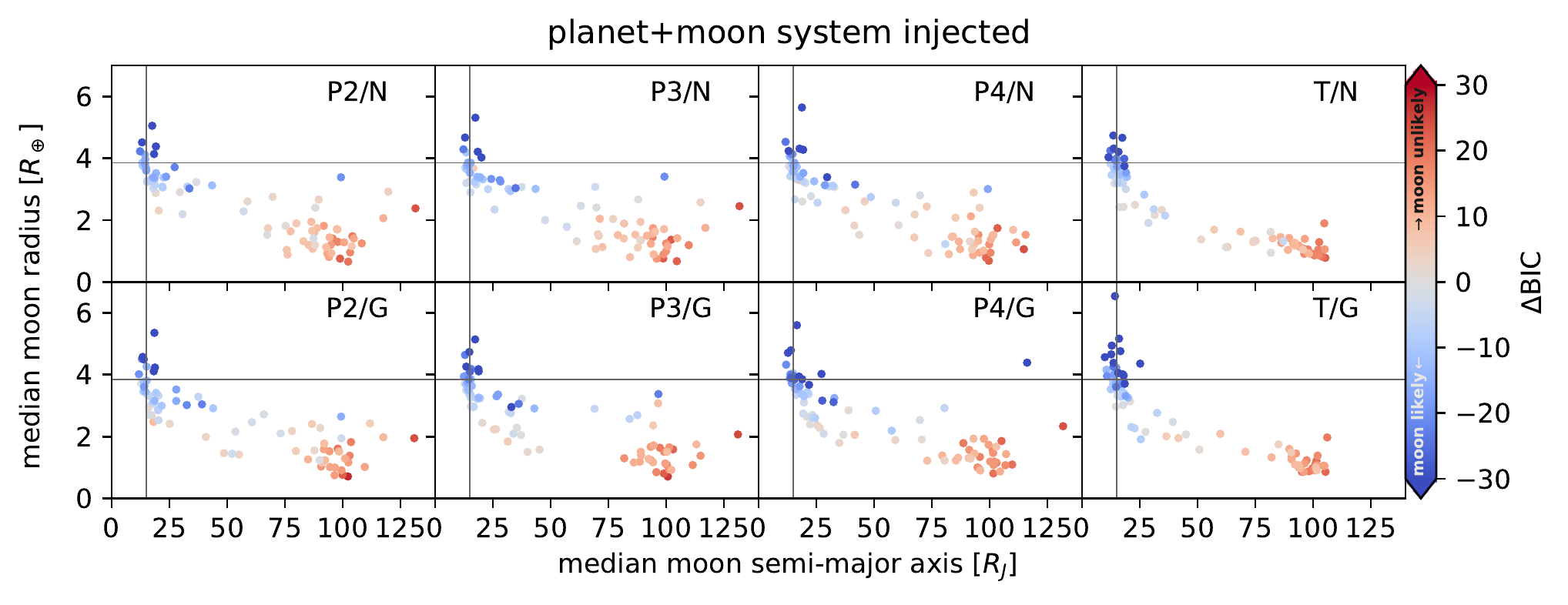}
\caption{Distribution of the median likelihood $R_\text{s}$ and $a_\text{s}$ for the transits injected into different parts of the Kepler-1625 light curve, using different detrending methods. The $\Delta{\rm BIC}$ of the planet-only model compared to the planet-moon model is indicated by the symbol color. The values of the moon semimajor axis (abscissa) and radius (ordinate) suggested by \cite{2018AJ....155...36T} are indicated with thin, gray lines in each sub-panel.}
\label{fig:injection_parameter_distribution}
\end{figure*}

In Fig.~\ref{fig:injection_delta_bic} we show the $\Delta$BIC for the 160 simulated Kepler light curves that were not rejected by our detrending method due to gaps very close to a transit. The left panel shows our results for the analysis of planet-only injections and the right panel refers to planet-moon injections. The tables in the panel headers list the true negative, false positive, true positive, and false positive rates as well as the rates of ambiguous cases. With ``positive'' (``negative''), we here refer to the detection (non-detection) of a moon.

In particular, we find the true negative rate (left panel, ${\Delta}{\rm BIC}~{\geq}~6$) to be between 65\,\% and 87.5\,\% and the true positive rate (right panel, ${\Delta}{\rm BIC}~{\leq}~-6$) to be between 31.25\,\% and 46.25\,\% depending on the detrending method, respectively.

The rates of false classifications is between 8.75\,\% and 17.5\,\% for the injected planet-only systems with a falsely detected moon (false positives) and between 30\,\% and 41.25\,\% for the injected planet-moon systems with a failed moon recovery (false negatives).

The rates of classification as a planet-moon system depends significantly on the treatment of gaps during the detrending procedure. Whenever the light curve is cut at a gap, the detection rates for a moon increase -- both for the false positives and for the true positives. Among all the detrending methods, this effect is especially strong for the trigonometric detrending. The false positive rate increases by almost a factor of two from 8.75\,\% (T/N) to 16.25\,\% (T/G) and the true positive rate increases by 15\,\% to 46.25\,\%. The effect on the true negative rate is strongest for the trigonometric detrending, decreasing from 87.5\,\% when the light curve is not cut at gaps (T/N) to 72.5\,\% if the light curve is cut (T/G). The false negative rate for the second order polynomial detrending decreases from 41.25\;\% (P2/N) to 30\,\% (P2/G) when gaps are cut, while the false negative rates of the other detrending methods remain almost unaffected.

Of all the light curves with an injected planet only, 21.25\,\% have an ambiguous classification with at least one of the detrending methods showing a negative and a different method showing a positive ${\Delta}{\rm BIC}$ above the threshold. For the light curves with an injected planet-moon system, there are 18.75\,\% with ambiguous classification and another 18.75\,\% of the injected planet-moon systems are classified unanimously as true positives by all detrending methods.

Fig.~\ref{fig:injection_parameter_distribution} shows the distribution of the retrieved moon parameters $a_\text{s}$ and $R_\text{s}$ as well as the corresponding ${\Delta}{\rm BIC}$ (see color scale) for each of the detrending methods.

For the light curves with an injected planet-moon system (lower set of panels), the maximum likelihood values of $a_\text{s}$ and $R_\text{s}$ of the true positives (blue) generally cluster around the injected parameters. In particular, we find that the moon turns out to be more likely (deeper-blue dots) when it is fitted to have a larger radius. The parameters of the false positives (blue dots in the upper set of panels) are more widely spread out, with moon radii ranging between 2 and 5~$R_\oplus$ and the moon semimajor axes spread out through essentially the entire parameter range that we explored. The clustering of median $a_{\rm s}$ at around $100~R_{\rm J}$ is an artifact of taking the median over a very unlocalized distribution along $a_{\rm s}$. For the polynomial detrending methods there are a certain number of what one could refer to as mischaracterized true positives. In these cases the ${\Delta}{\rm BIC}$-based planet-only vs. planet-moon classification is correct but the maximum likelihood values are very different from the injected ones.

The correctly identified planet-only systems show a similar distribution of $a_\text{s}$ and $R_\text{s}$ as in our experiment with white noise only and a 700\,ppm amplitude (Fig.~\ref{fig:white_noise_delta_bic}). 

Most surprisingly, and potentially most worryingly, the false positives (blue dots in the upper set of panels in Fig.~\ref{fig:injection_parameter_distribution}) cluster around the values of the moon parameters found by \cite{2018AJ....155...36T}, in particular if the light curve is cut at the first gap.

\section{Discussion}

In this article we compare several detrending methods of the light curve of Kepler-1625, some of which were used by \citet{2018AJ....155...36T} in their characterization of the exomoon candidate around Kepler-1625~b. However, we do not perform an exhaustive survey of all available detrending methods, such as Gaussian processes \citep{2016MNRAS.459.2408A}.

We show that the sequential detrending and fitting procedure of transit light curves is prone to introducing features that can be misinterpreted as signal, in our case as an exomoon. This ``pre-whitening'' method of the data has thus to be used with caution. Our investigations of a polynomial-based fitting and of a trigonometric detrending procedure show that the resulting best-fit model depends strongly on the specific detrending function, e.g. on the order of the polynomial or on the minimum time scale (or wavelength) of a cosine. This is crucial for any search of secondary effects in the transit light curves -- moons, rings, evaporating atmospheres etc. -- and is in stark contrast to a claim by \citet{2017AJ....153..193A}, who stated that neither the choice of the detrending function nor the choice of the detrending window of the light curve would have a significant effect on the result. We find that this might be true on a by-eye level but not on a level of 100\,ppm or below. Part of the difference between our findings and those of \citet{2017AJ....153..193A} could be in the different time scales we investigate. While they considered the effect of stellar flairs on time scales of less than a day, much less than the $\sim$2 day transit duration of their specific target, our procedure operates on various time  scales of up to several weeks. Moreover, we develop a dynamical moon model to fit multiple transits, whereas \citet{2017AJ....153..193A} study only a single transit.

Since the actual presence and the putative orbital position of a hypothetical exomoon around Kepler-1625\,b is unknown a priori, it is unclear how much of the light curve would need to be protected from (or neglected for) the pre-fit detrending process in order to avoid a detrending of a possible moon signal itself. In turn, we show that in the case of Kepler-1625\, different choices for this protected time scale around the transit yield different confidences and different solutions for a planet-moon system. We find that the previously announced solution by \citet{2018AJ....155...36T} is only one of many possibilities with similar likelihoods (specifically: Bayesian Information Criteria). This suggests, but by no means proves, that all of these solutions could, in fact, be due to red noise artifacts (e.g. stellar or instrumental) rather than indicative of a moon signal.

Our finding of higher true positive rate compared to a false positive rate from injection-retrieval experiments could be interpreted as slight evidence in favor of a genuine exomoon. This interpretation, however, depends on the number of transiting planets and planet candidates around stars with similar noise characteristics that were included in the \cite{2018AJ....155...36T} search. Broadly speaking, if more than a handful of similar targets were studied, the probability of at least one false positive detection becomes quite likely.

\section{Conclusions}

We investigated the detrending of the transit light curve of Kepler-1625\,b with a method very similar to the one used by \cite{2018AJ....155...36T} and then applied a Bayesian framework with MCMC modeling to search for a moon. Our finding of a $\Delta$BIC of $-4.954$ favors the planet-moon over the planet-only hypothesis. Although significant, this tentative detection fails to cross the threshold of $-6$, which we would consider strong evidence of a moon. Our $\Delta$BIC value would certainly change if we could include the additional data from the high-precision transit observations executed in October 2017 with the Hubble Space Telescope \citep{2018AJ....155...36T} in our analysis. Moreover, by varying the free parameters of our detrending procedure, we also find completely different solutions for a planet-moon system, i.e. different planet-moon orbital configurations during transits and different moon radii or planet-moon orbital semimajor axes.

As an extension to this validation of the previously published work, we performed 200 injection-retrieval experiments into the original out-of-transit parts of the Kepler light curve. We also extended the previous work by exploring different detrending methods, such as second-, third-, and fourth-order polynomials as well as trigonometric methods and find false-positive rates between 8.75\,\% and 16.25\,\%, depending on the method. Surprisingly, we find that the moon radius and planet-moon distances of these false positives are very similar to the ones measured by \cite{2018AJ....155...36T}. In other words, in 8.75\,\% to 16.25\,\% of the light curves that contained an artificially injected planet only, we find a moon that is about as large as Neptune and orbits Kepler-1625\,b at about $20\,R_{\rm J}$.

To sum up, we find tentative statistical evidence for a moon in this particular Kepler light curve of Kepler-1625, but we also show that the significant fraction of similar light curves, which contained a planet only, would nevertheless indicate a moon with properties similar to the candidate Kepler-1625\,b-i. Clearly, stellar and systematic red noise components are the ultimate barrier to an unambiguous exomoon detection around Kepler-1625\,b and follow-up observations have the potential of solving this riddle based on the framework that we present.

Of all the detrending methods we investigated, the trigonometric method, which is very similar to the CoFiAM method of \citet{2018AJ....155...36T}, can produce the highest true positive rate. At the same time, however, this method also ranks among the ones producing the highest false positive rates as well. To conclude, we recommend that any future exomoon candidate be detrended with as many different detrending methods as possible to evaluate the robustness of the classification.

\begin{acknowledgements}
We thank James Kuszlewicz and Jesper Schou for useful discussions. This work was supported in part by the German Aerospace Center (DLR) under PLATO Data Center grant 50OL1701. This paper includes data collected by the Kepler mission. Funding for the Kepler mission is provided by the NASA Science Mission directorate. This work has made use of data provided by NASA and the Space Telescope Science Institute.  K.R. is a member of the International Max Planck Research School for Solar System Science at the University of G\"ottingen. K.R. contributed to the analysis of the simulated light curves, to the interpretation of the results, and to the writing of the article. 
\end{acknowledgements}

\bibliographystyle{aa} 
\bibliography{ms}

\begin{thebibliography}{45}
\expandafter\ifx\csname natexlab\endcsname\relax\def\natexlab#1{#1}\fi

\bibitem[{{Agol} {et~al.}(2015){Agol}, {Jansen}, {Lacy}, {Robinson}, \&
  {Meadows}}]{2015ApJ...812....5A}
{Agol}, E., {Jansen}, T., {Lacy}, B., {Robinson}, T.~D., \& {Meadows}, V. 2015,
  \apj, 812, 5

\bibitem[{{Aigrain} {et~al.}(2015){Aigrain}, {Hodgkin}, {Irwin}, {Lewis}, \&
  {Roberts}}]{2015MNRAS.447.2880A}
{Aigrain}, S., {Hodgkin}, S.~T., {Irwin}, M.~J., {Lewis}, J.~R., \& {Roberts},
  S.~J. 2015, \mnras, 447, 2880

\bibitem[{{Aigrain} \& {Irwin}(2004)}]{2004MNRAS.350..331A}
{Aigrain}, S. \& {Irwin}, M. 2004, \mnras, 350, 331

\bibitem[{{Aigrain} {et~al.}(2016){Aigrain}, {Parviainen}, \&
  {Pope}}]{2016MNRAS.459.2408A}
{Aigrain}, S., {Parviainen}, H., \& {Pope}, B.~J.~S. 2016, \mnras, 459, 2408

\bibitem[{{Aizawa} {et~al.}(2017){Aizawa}, {Uehara}, {Masuda}, {Kawahara}, \&
  {Suto}}]{2017AJ....153..193A}
{Aizawa}, M., {Uehara}, S., {Masuda}, K., {Kawahara}, H., \& {Suto}, Y. 2017,
  \aj, 153, 193

\bibitem[{Ben-Jaffel \& Ballester(2014)}]{BenJaffel2014}
Ben-Jaffel, L. \& Ballester, G.~E. 2014, The Astrophysical Journal Letters,
  785, L30

\bibitem[{{Cabrera, J.} \& {Schneider, J.}(2007)}]{Cabrera2007}
{Cabrera, J.} \& {Schneider, J.} 2007, A\&A, 464, 1133

\bibitem[{{Domingos} {et~al.}(2006){Domingos}, {Winter}, \&
  {Yokoyama}}]{2006MNRAS.373.1227D}
{Domingos}, R.~C., {Winter}, O.~C., \& {Yokoyama}, T. 2006, \mnras, 373, 1227

\bibitem[{{Dumusque} {et~al.}(2012){Dumusque}, {Pepe}, {Lovis},
  {S{\'e}gransan}, {Sahlmann}, {Benz}, {Bouchy}, {Mayor}, {Queloz}, {Santos},
  \& {Udry}}]{2012Natur.491..207D}
{Dumusque}, X., {Pepe}, F., {Lovis}, C., {et~al.} 2012, \nat, 491, 207

\bibitem[{Durbin \& Watson(1950)}]{10.2307/2332391}
Durbin, J. \& Watson, G.~S. 1950, Biometrika, 37, 409

\bibitem[{{Foreman-Mackey} {et~al.}(2013){Foreman-Mackey}, {Hogg}, {Lang}, \&
  {Goodman}}]{2013PASP..125..306F}
{Foreman-Mackey}, D., {Hogg}, D.~W., {Lang}, D., \& {Goodman}, J. 2013, \pasp,
  125, 306

\bibitem[{Forgan(2017)}]{Forgan2017}
Forgan, D.~H. 2017, Monthly Notices of the Royal Astronomical Society, 470, 416

\bibitem[{{Gibson} {et~al.}(2012){Gibson}, {Aigrain}, {Roberts}, {Evans},
  {Osborne}, \& {Pont}}]{2012MNRAS.419.2683G}
{Gibson}, N.~P., {Aigrain}, S., {Roberts}, S., {et~al.} 2012, \mnras, 419, 2683

\bibitem[{{Han} \& {Han}(2002)}]{2002ApJ...580..490H}
{Han}, C. \& {Han}, W. 2002, \apj, 580, 490

\bibitem[{{Heller}(2014)}]{2014ApJ...787...14H}
{Heller}, R. 2014, \apj, 787, 14

\bibitem[{Heller(2017)}]{HellerReview2017}
Heller, R. 2017, Detecting and Characterizing Exomoons and Exorings, ed. H.~J.
  Deeg \& J.~A. Belmonte (Cham: Springer International Publishing), 1--17

\bibitem[{{Heller}(2018)}]{2018A&A...610A..39H}
{Heller}, R. 2018, \aap, 610, A39

\bibitem[{{Heller} \& {Albrecht}(2014)}]{2014ApJ...796L...1H}
{Heller}, R. \& {Albrecht}, S. 2014, \apjl, 796, L1

\bibitem[{{Heller} {et~al.}(2016{\natexlab{a}}){Heller}, {Hippke}, \&
  {Jackson}}]{2016ApJ...820...88H}
{Heller}, R., {Hippke}, M., \& {Jackson}, B. 2016{\natexlab{a}}, \apj, 820, 88

\bibitem[{{Heller} {et~al.}(2016{\natexlab{b}}){Heller}, {Hippke}, {Placek},
  {Angerhausen}, \& {Agol}}]{2016A&A...591A..67H}
{Heller}, R., {Hippke}, M., {Placek}, B., {Angerhausen}, D., \& {Agol}, E.
  2016{\natexlab{b}}, \aap, 591, A67

\bibitem[{{Heller} {et~al.}(2014){Heller}, {Williams}, {Kipping}, {Limbach},
  {Turner}, {Greenberg}, {Sasaki}, {Bolmont}, {Grasset}, {Lewis}, {Barnes}, \&
  {Zuluaga}}]{2014AsBio..14..798H}
{Heller}, R., {Williams}, D., {Kipping}, D., {et~al.} 2014, Astrobiology, 14,
  798

\bibitem[{{Hippke}(2015)}]{2015ApJ...806...51H}
{Hippke}, M. 2015, \apj, 806, 51

\bibitem[{Hippke(2018)}]{hippke_zenodo}
Hippke, M. 2018, Synthetic Dataset For The Exomoon Candidate Around Kepler-1625
  b

\bibitem[{Kass \& Raftery(1995)}]{kass_1995}
Kass, R.~E. \& Raftery, A.~E. 1995, Journal of the American Statistical
  Association, 90, 773

\bibitem[{{Kipping}(2009)}]{2009MNRAS.392..181K}
{Kipping}, D.~M. 2009, \mnras, 392, 181

\bibitem[{{Kipping}(2013)}]{2013MNRAS.435.2152K}
{Kipping}, D.~M. 2013, \mnras, 435, 2152

\bibitem[{{Kipping} {et~al.}(2012){Kipping}, {Bakos}, {Buchhave},
  {Nesvorn{\'y}}, \& {Schmitt}}]{2012ApJ...750..115K}
{Kipping}, D.~M., {Bakos}, G.~{\'A}., {Buchhave}, L., {Nesvorn{\'y}}, D., \&
  {Schmitt}, A. 2012, \apj, 750, 115

\bibitem[{{Kipping} {et~al.}(2013{\natexlab{a}}){Kipping}, {Forgan}, {Hartman},
  {Nesvorn{\'y}}, {Bakos}, {Schmitt}, \& {Buchhave}}]{2013ApJ...777..134K}
{Kipping}, D.~M., {Forgan}, D., {Hartman}, J., {et~al.} 2013{\natexlab{a}},
  \apj, 777, 134

\bibitem[{{Kipping} {et~al.}(2013{\natexlab{b}}){Kipping}, {Hartman},
  {Buchhave}, {Schmitt}, {Bakos}, \& {Nesvorn{\'y}}}]{2013ApJ...770..101K}
{Kipping}, D.~M., {Hartman}, J., {Buchhave}, L.~A., {et~al.}
  2013{\natexlab{b}}, \apj, 770, 101

\bibitem[{{Kipping} {et~al.}(2014){Kipping}, {Nesvorn{\'y}}, {Buchhave},
  {Hartman}, {Bakos}, \& {Schmitt}}]{2014ApJ...784...28K}
{Kipping}, D.~M., {Nesvorn{\'y}}, D., {Buchhave}, L.~A., {et~al.} 2014, \apj,
  784, 28

\bibitem[{{Kipping} {et~al.}(2015){Kipping}, {Schmitt}, {Huang}, {Torres},
  {Nesvorn{\'y}}, {Buchhave}, {Hartman}, \& {Bakos}}]{2015ApJ...813...14K}
{Kipping}, D.~M., {Schmitt}, A.~R., {Huang}, X., {et~al.} 2015, \apj, 813, 14

\bibitem[{{Lecavelier des Etangs} {et~al.}(2017){Lecavelier des Etangs},
  {H{\'e}brard}, {Blandin}, {Cassier}, {Deeg}, {Bonomo}, {Bouchy},
  {D{\'e}sert}, {Ehrenreich}, {Deleuil}, {D{\'{\i}}az}, {Moutou}, \&
  {Vidal-Madjar}}]{2017A&A...603A.115L}
{Lecavelier des Etangs}, A., {H{\'e}brard}, G., {Blandin}, S., {et~al.} 2017,
  \aap, 603, A115

\bibitem[{{Lewis} \& {Fujii}(2014)}]{2014ApJ...791L..26L}
{Lewis}, K.~M. \& {Fujii}, Y. 2014, \apjl, 791, L26

\bibitem[{{Mandel} \& {Agol}(2002)}]{2002ApJ...580L.171M}
{Mandel}, K. \& {Agol}, E. 2002, \apjl, 580, L171

\bibitem[{{Mathur} {et~al.}(2017){Mathur}, {Huber}, {Batalha}, {Ciardi},
  {Bastien}, {Bieryla}, {Buchhave}, {Cochran}, {Endl}, {Esquerdo}, {Furlan},
  {Howard}, {Howell}, {Isaacson}, {Latham}, {MacQueen}, \&
  {Silva}}]{2017ApJS..229...30M}
{Mathur}, S., {Huber}, D., {Batalha}, N.~M., {et~al.} 2017, \apjs, 229, 30

\bibitem[{{Moskovitz} {et~al.}(2009){Moskovitz}, {Gaidos}, \&
  {Williams}}]{2009AsBio...9..269M}
{Moskovitz}, N.~A., {Gaidos}, E., \& {Williams}, D.~M. 2009, Astrobiology, 9,
  269

\bibitem[{Peters \& Turner(2013)}]{Peters2013}
Peters, M.~A. \& Turner, E.~L. 2013, The Astrophysical Journal, 769, 98

\bibitem[{{Rajpaul} {et~al.}(2016){Rajpaul}, {Aigrain}, \&
  {Roberts}}]{2016MNRAS.456L...6R}
{Rajpaul}, V., {Aigrain}, S., \& {Roberts}, S. 2016, \mnras, 456, L6

\bibitem[{{Sartoretti} \& {Schneider}(1999)}]{1999A&AS..134..553S}
{Sartoretti}, P. \& {Schneider}, J. 1999, \aaps, 134, 553

\bibitem[{{Schwarz}(1978)}]{1978AnSta...6..461S}
{Schwarz}, G. 1978, Annals of Statistics, 6, 461

\bibitem[{{Seager} \& {Mall{\'e}n-Ornelas}(2003)}]{2003ApJ...585.1038S}
{Seager}, S. \& {Mall{\'e}n-Ornelas}, G. 2003, \apj, 585, 1038

\bibitem[{{Simon} {et~al.}(2010){Simon}, {Szab{\'o}}, {Szatm{\'a}ry}, \&
  {Kiss}}]{2010MNRAS.406.2038S}
{Simon}, A.~E., {Szab{\'o}}, G.~M., {Szatm{\'a}ry}, K., \& {Kiss}, L.~L. 2010,
  \mnras, 406, 2038

\bibitem[{{Szab{\'o}} {et~al.}(2013){Szab{\'o}}, {Szab{\'o}}, {D{\'a}lya},
  {Simon}, {Hodos{\'a}n}, \& {Kiss}}]{2013A&A...553A..17S}
{Szab{\'o}}, R., {Szab{\'o}}, G., {D{\'a}lya}, G., {et~al.} 2013, \aap, 553,
  A17

\bibitem[{{Teachey} {et~al.}(2018){Teachey}, {Kipping}, \&
  {Schmitt}}]{2018AJ....155...36T}
{Teachey}, A., {Kipping}, D.~M., \& {Schmitt}, A.~R. 2018, \aj, 155, 36

\bibitem[{{Vanderburg} {et~al.}(2018){Vanderburg}, {Rappaport}, \&
  {Mayo}}]{2018arXiv180501903V}
{Vanderburg}, A., {Rappaport}, S.~A., \& {Mayo}, A.~W. 2018, ArXiv e-prints
  [\eprint[arXiv]{1805.01903}]

\end{thebibliography}


\begin{appendix} 
\section{Effect of the window length on the Bayesian Information Criterion}
\label{sec:appendix_Bayesian}

Given the constraint of orbital stability, a moon can only possibly orbit its planet within the planet's Hill sphere. Hence, transits may only occur within a certain time interval around the midpoint of the planetary transit. This time $t_\text{Hill}$ can be calculated as
\begin{align}
t_\text{Hill}=\frac{\eta \, R_\text{Hill}}{v_\text{orbit}}=\eta\frac{P}{2\pi}\sqrt{\frac{M_\text{p}}{3M_\star}} \ \ ,
\end{align}
\noindent
where $v_\text{orbit}$ is the orbital velocity of the planet-moon system around the star, $M_\text{p}$ and $M_\star$ the planet and star mass, $P$ the orbital period of the planet-moon system, and $R_{\rm Hill}$ is the Hill radius of the planet. $\eta$ is a factor between 0 and 1, which has been numerically determined for prograde moons ($\eta~\approx~0.5$) and for retrograde moons ($\eta~\approx~1$), details depending on the orbital eccentricities \citep{2006MNRAS.373.1227D}. We focus on prograde moons and choose $\eta=0.5$. For a $10~M_{\rm J}$ planet in a $287$~d orbit around a $1.1~M_\odot$ star the Hill time is $t_\text{Hill}=3.25~$d.

\begin{figure*}[h!]
\centering
\includegraphics[width=0.495\linewidth]{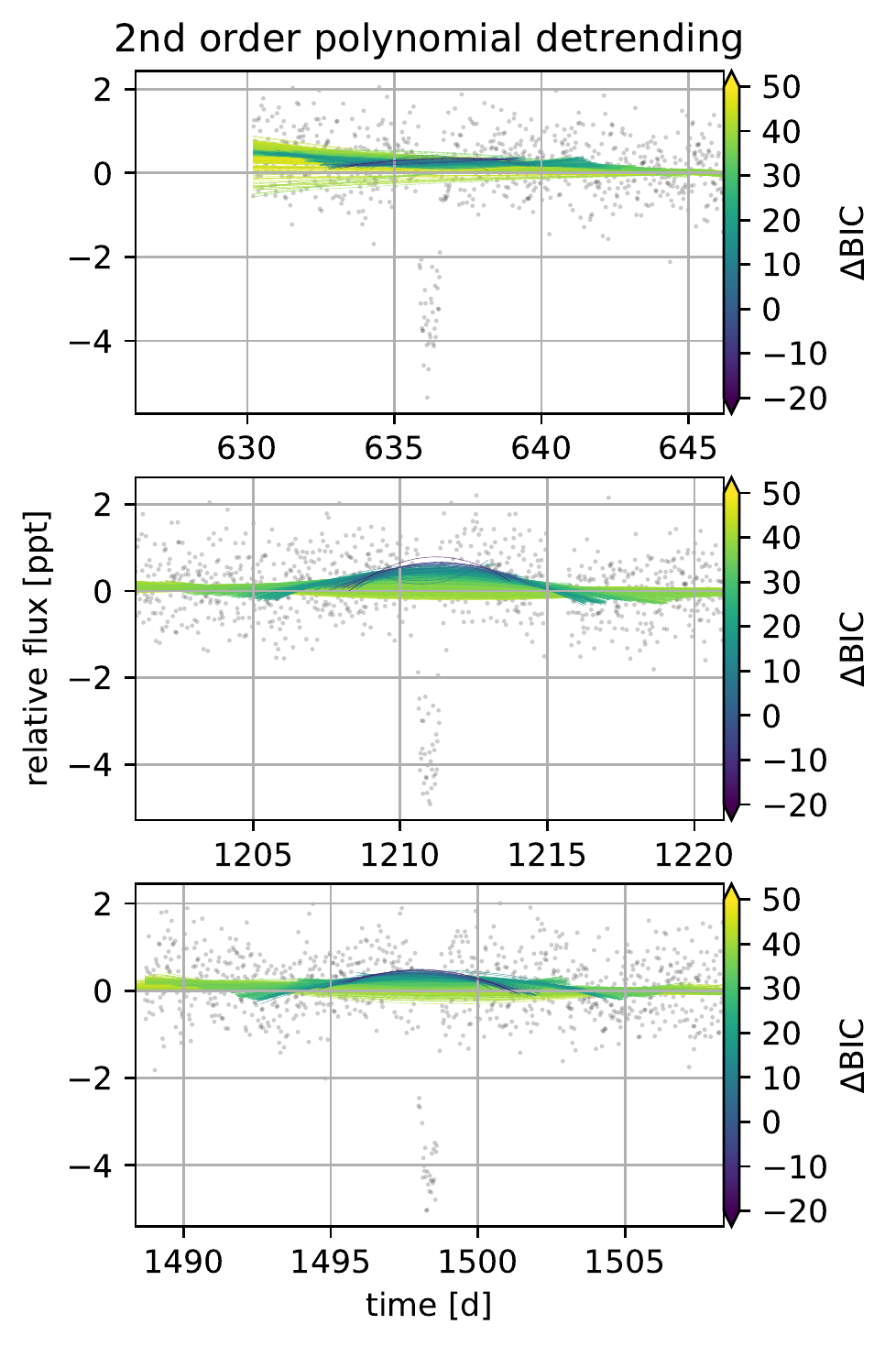}
\includegraphics[width=0.495\linewidth]{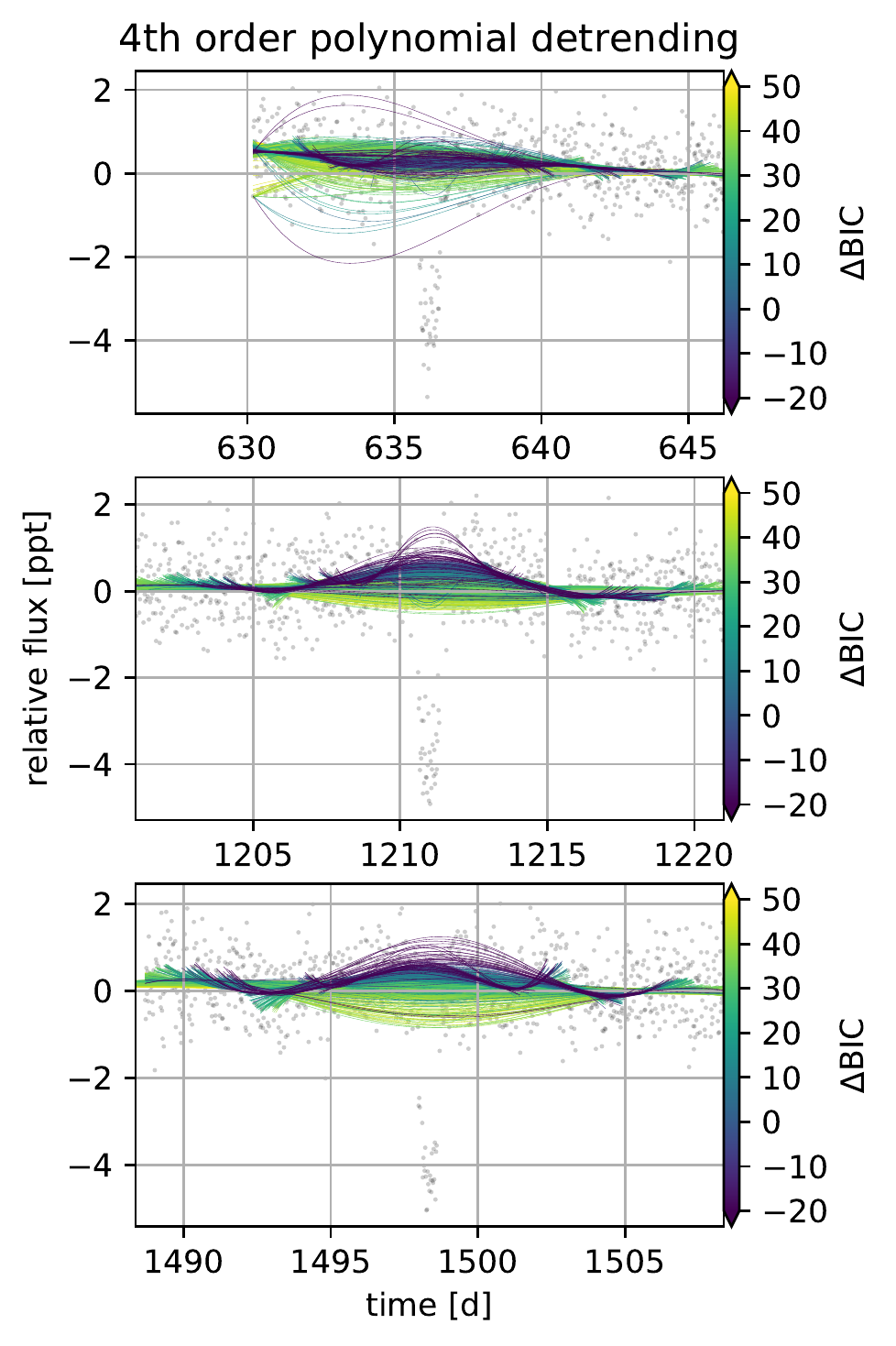}
\caption{Detrending for different cutout times $t_{\rm c}$ and base length $D$, color coded by the resulting $\Delta$BIC using a 2nd- and 4th-order polynomial function. While some of the detrending models corresponding to a large negative $\Delta$BIC are clearly results of wrong detrending, it is much less clear for many other detrending models.}
\label{fig:cutwidth}
\end{figure*}

As shown in Fig.~\ref{fig:detrending}, the length of the light curve, which is neglected for the polynomial fit has a strong effect on the resulting detrended light curve. Figure~\ref{fig:cutwidth} shows the effect that different cutout times $t_c$ and detrending base lines $D$ can have on whether a moon is detected or not.
\clearpage
\section{Autocorrelation of Detrended Light Curves}
\label{sec:appendix_autocorrelation}

\begin{figure}
\centering
\includegraphics[width=0.495\textwidth]{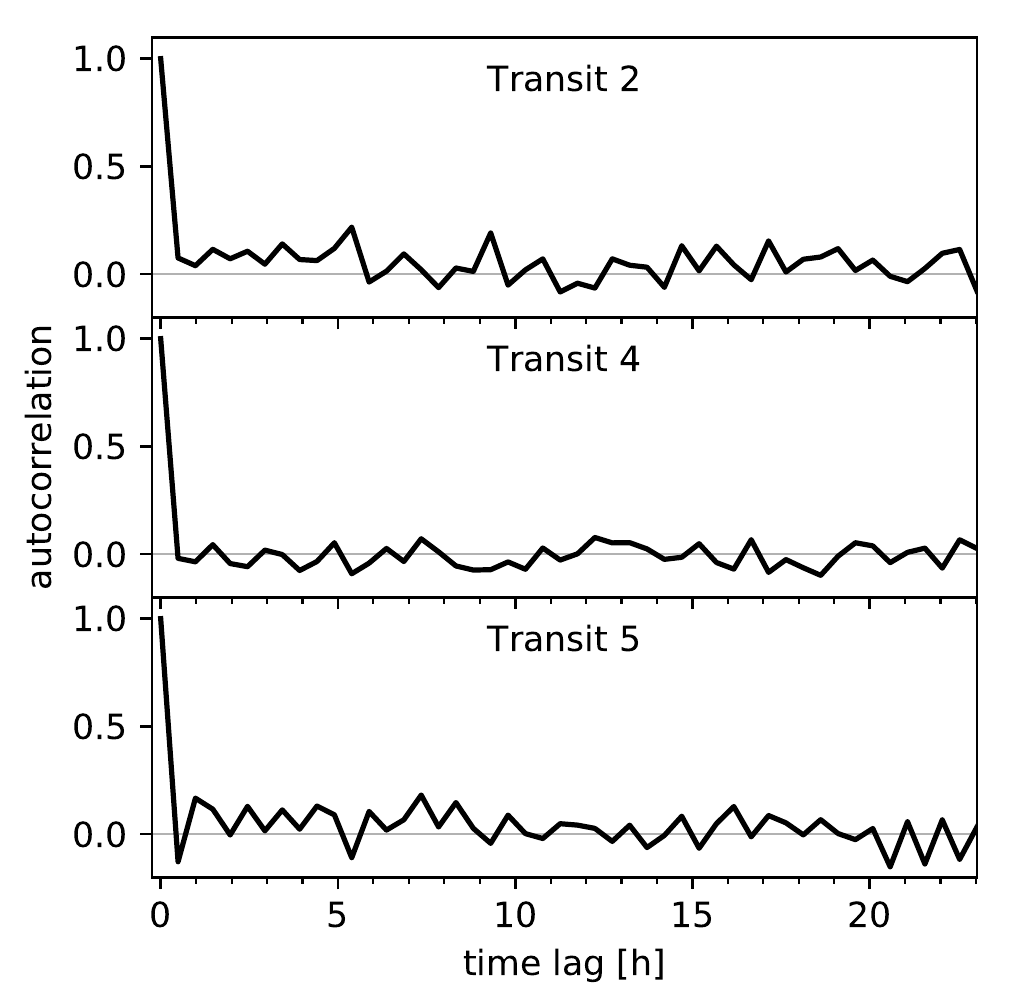}
\caption{The autocorrelation of the difference between the detrended light curve and the best fitting model.}
\label{fig:autocorrelation}
\end{figure}

The autocorrelations of the detrended light curves are shown in Fig.~\ref{fig:autocorrelation}.
For all three transits, the autocorrelation is close to zero, except for the zero-lag component. This suggests that it is reasonable to model the noise covariance matrix as a diagonal matrix (see Sect.~\ref{sec:posterior}).
\end{appendix}
\end{document}